\documentclass[eqsecnum,twocolumn,showpacs,amsmath,amssymb,bibnotes]{revtex4}
\usepackage{bm}
\usepackage{graphicx,psfig,epsfig,subfigure,afterpage}
\usepackage{amsfonts}

\newcommand{\djs}{{\mathfrak{D}}_{\rm M}}
\newcommand{\df}{{\mathfrak{D}}_{\rm F}}
\newcommand{\di}{{\mathfrak{D}}_{\rm i}}

\newcommand{\tdjs}{{\mathfrak{\tilde{D}}}_{\rm M}}
\newcommand{\tdf}{{\mathfrak{\tilde{D}}}_{\rm F}}
\newcommand{\tdi}{{\mathfrak{\tilde{D}}}_{\rm i}}

\newcommand{\hdjs}{{\mathfrak{\hat{D}}}_{\rm M}}

\newcommand{\vect}[1]{\underline{#1}}
\newcommand{\tens}[1]{\underline{\underline{#1}}}

\newcommand{\vm}{\vect{v}_{\,\rm m}}
\newcommand{\vs}{\vect{v}_{\,\rm s}}

\newcommand{\vrel}{\vect{v}_{\, \rm rel}}
\newcommand{\vv}{\vect{v}}

\newcommand{\etam}{\eta_{\,\rm m}}
\newcommand{\etas}{\eta_{\,\rm s}}

\newcommand{\Dm}{\tens{D}_{\,\rm m}}
\newcommand{\Ds}{\tens{D}_{\,\rm s}}

\newcommand{\Drel}{\tens{D}_{\,\rm rel}}
\newcommand{\Dv}{\tens{D}}

\newcommand{\Omm}{\tens{\Omega}_{\,\rm m}}

\newcommand{\nablu}{\vect{\nabla}}

\newcommand{\be}{\begin{equation}}
\newcommand{\ee}{\end{equation}}
\newcommand{\bea}{\begin{eqnarray}}
\newcommand{\eea}{\end{eqnarray}}

\newcommand{\gdot}{\dot{\gamma}}
\newcommand{\bgdot}{\bar{\dot{\gamma}}}

\newcommand{\D}{\displaystyle}

\newcommand{\gae}{\stackrel{>}{\scriptstyle\sim}}
\newcommand{\lae}{\stackrel{<}{\scriptstyle\sim}}
\newcommand{\ie}{{\it i.e.\/}}
\newcommand{\eg}{{\it e.g.\/}}

\newcommand{\versus}{{\it vs.\/}}

\newcommand{\etab}{{\eta}}
\newcommand{\bw}{\begin{widetext}}
\newcommand{\ew}{\end{widetext}}

\newcommand{\bmini}{\begin{minipage}}
\newcommand{\emini}{\end{minipage}}

\newcommand{\eigenvec}{\vect{{\tt{v}}}_{\vect{k},\alpha}}

\begin{document}

\title{Kinetics of the shear banding instability in startup flows}
\author{S. M. Fielding}
\email{physf@irc.leeds.ac.uk} 
\author{P. D. Olmsted}
\email{p.d.olmsted@leeds.ac.uk}
\affiliation{Polymer IRC and
  Department of Physics \& Astronomy, University of Leeds, Leeds LS2
  9JT, United Kingdom} 
\date{\today} 
\begin{abstract}
  Motivated by recent experiments on semi-dilute wormlike micelles, we
  study the early stages of the shear banding instability using the
  non-local Johnson-Segalman model with a ``two-fluid'' coupling to
  concentration. We perform a linear stability analysis for coupled
  fluctuations in shear rate $\gdot$, micellar strain $\tens{W}$ and
  concentration $\phi$ about an initially homogeneous state.  This
  resembles the Cahn-Hilliard (CH) analysis of fluid-fluid demixing,
  though we discuss important differences.  First assuming the
  homogeneous state to lie on the intrinsic constitutive curve, we
  calculate the ``spinodal'' onset of instability in sweeps along this
  constitutive curve. We then consider startup ``quenches'' into the
  unstable region. Here the instability in general occurs before the
  intrinsic constitutive curve can be attained so we analyse the
  fluctuations with respect to the {\em time-dependent} homogeneous
  startup flow, to find the selected length and time scales at which
  inhomogeneity first emerges. In the uncoupled limit, fluctuations in
  $\gdot$ and $\tens{W}$ are independent of those in $\phi$, and are
  unstable when the intrinsic constitutive curve has negative slope;
  but no length scale is selected.  For the coupled case, this
  instability is enhanced at short length scales via feedback with
  $\phi$ and a length scale is selected, consistently with the recent
  experiments. The unstable region is then broadened by an extent that
  increases with proximity to an underlying (zero-shear) CH demixing
  instability. Far from demixing, the broadening is slight and the
  instability is still dominated by $\delta\gdot$ and $\delta
  \tens{W}$ with only small $ \delta\phi$.  Close to demixing,
  instability sets in at very low shear rate, where it is demixing
  triggered by flow.

\end{abstract}
\pacs{{47.50.+d}{ Non-Newtonian fluid flows}--
     {47.20.-k}{ Hydrodynamic stability}--
     {36.20.-r}{ Macromolecules and polymer molecules}
}
\maketitle
\vskip10truept

\section{Introduction}
\label{sec:intro}

\begin{figure}[h]
\begin{center}
 \centerline{\psfig{figure=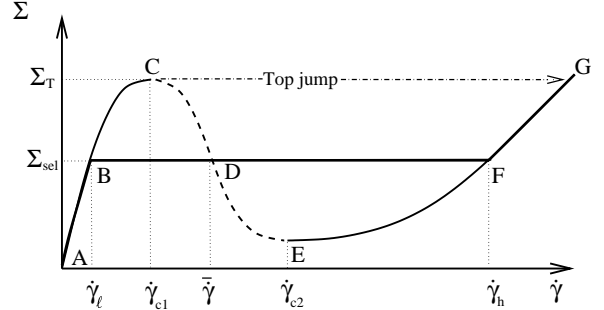,width=8cm}}
\caption{Schematic flow curve for  wormlike micelles.
\label{fig:schem} } 
\end{center}
\end{figure}
\begin{figure}[h]
  \centerline{\psfig{figure=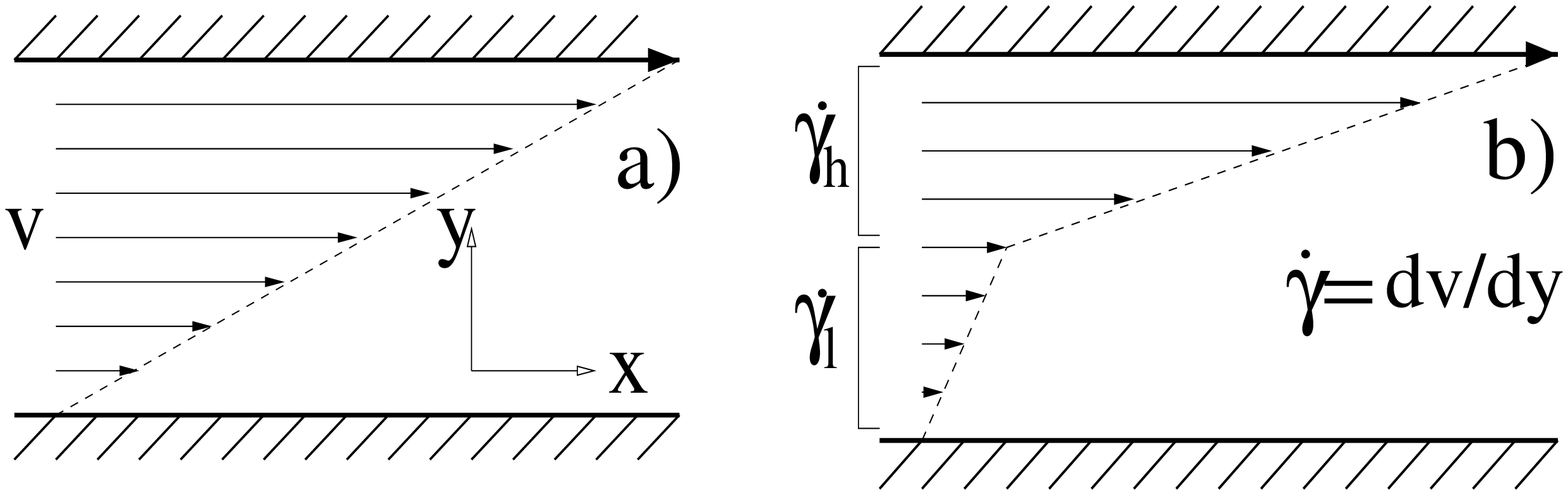,width=8cm}}
\caption{(a) Homogeneous shear rate and (b) banded profiles.
\label{fig:picture} } 
\end{figure}

For many complex fluids, the intrinsic constitutive curve of shear
stress $\Sigma$ as a function of shear rate $\gdot$ is non-monotonic,
admitting multiple values of shear rate at common stress. For example,
Cates' model for semi-dilute wormlike micelles~\cite{cates90} predicts
that the steady shear stress decrease above a critical
$\gdot=\gdot_{\rm c1}$.  At very high shear rates, fast relaxation
processes must eventually restore an increasing
stress~\cite{SpenCate94,SCM93}. See curve ACEG of
Fig.~\ref{fig:schem}.  For the range $\gdot_{\rm c1}<\gdot<\gdot_{\rm
  c2}$ in which the stress decreases, steady homogeneous flow
(Fig.~\ref{fig:picture}a) is unstable~\cite{Yerushalmi70}. For an
applied shear rate $\bar{\gdot}$ in this unstable range, Spenley,
Cates and McLeish~\cite{SCM93} proposed that the system separates into
high and low shear rate bands ($\gdot_{\rm h}$ and $\gdot_\ell$) with
relative volume fractions satisfying the applied shear rate
$\bar{\gdot}$.  (Fig.~\ref{fig:picture}b.)  The steady state flow
curve then has the form ABFG. Within the banding regime, BF, a change
in the applied shear rate adjusts the relative fraction of the bands,
while the stress $\Sigma_{\rm sel}$ (common to both) remains constant.
Several constitutive models augmented with interfacial gradient terms
have captured this
behaviour-~\cite{olmsted99a,lu99,olmstedlu97,olmsted92,Pear94,Yuan99b,spenley96}.


Experimentally, this scenario is well established for shear-thinning
wormlike micelles~\cite{berret94b,Call+96,grand97}. The steady state
flow curve has a well defined, reproducible plateau $\Sigma_{\rm
  sel}$.  Coexistence of high and low viscosity bands has been
observed by NMR
spectroscopy~\cite{Call+96,MairCall96,MairCall96c,BritCall97}. Further
evidence comes from small angle neutron
scattering~\cite{Berr+94,schmitt94,Capp+97,BRL98,rehage91,berret94b};
and from flow birefringence (FB)~\cite{Decr+95,Makh+95,DCC97,BPD97},
which reveals a (quasi) nematic birefringence band coexisting with an
isotropic one.  The nematic band of FB has commonly been identified
with the low viscosity band of NMR; but see~\cite{FisCal01,FisCal00}.


Here we consider banding formation kinetics.
Experimentally~\cite{berret94b,Berr97,grand97,decruppe98,DecLerBer01,LerDecBer00,BerrPort99},
in rapid upward stress sweeps the shear rate initially follows the
steady state flow curve (AB in Fig.~\ref{fig:schem}) before departing
for stresses $\Sigma>\Sigma_{\rm sel}$ along a metastable branch (BC).
When  this branch starts to level off (hinting of an unstable
branch for $\gdot\gae\gdot_{\rm c1}$) the shear rate finally
``top-jumps''. Under shear startup in the metastable region
$\gdot_{\ell}<\gdot\lae\gdot_{\rm c1}$, the stress first rapidly attains
the metastable branch BC (sometimes via oscillations), before slowly
decaying onto the steady-state plateau $\Sigma_{\rm sel}$ via a
``sigmoidal'' envelope $\exp[-(t/\tau_{\rm NG})^\alpha]$, with
stretching exponent $\alpha$.  The time scale $\tau_{\rm NG}$ greatly
exceeds the Maxwell time $\tau$ of linear rheology, but is smaller for
shear rates $\gdot$ closer to $\gdot_{\rm c1}$; for most systems the
stretching exponent is in the range $1.5\lae\alpha\lae 3$. In the data
of \cite{BerrPort99}, for example, $\tau_{\rm NG}\sim
(\gdot-\gdot_{\ell})^{-p}$ and $\alpha\approx 2$ throughout the
metastable range, with a crossover to $\alpha\approx 1$ for
$\gdot\gae\gdot_{\rm c1}$ signifying onset of instability. In more
dilute systems~\cite{DecLerBer01} the onset of instability can be
marked by a huge stress overshoot subsiding rapidly to $\Sigma_{\rm
  sel}$ via damped oscillations.

In the same experiments~\cite{DecLerBer01} the stress overshoot
coincided with pronounced concentration fluctuations that first
emerged perpendicular to the shear compression axis, at a selected
length scale $O(1\mu \rm{m})$.  These fluctuations were attributed by
the authors of Ref.~\cite{DecLerBer01} to the
Helfand-Fredrickson~\cite{HelfFred89} coupling of concentration to
flow.  Although this mechanism has not, to date, been widely employed
in the theory of shear banding -- an important exception
being~\cite{schmitt95} (discussed below) -- it has been recognised
elsewhere~\cite{brochdgen77,HelfFred89,DoiOnuk92,milner93,WPD91,ClarMcle98,beris94,SBJ97,tanak96}
as an important feature of two components systems with widely
separated relaxation times (\eg\ polymer and solvent).  The slow
component (polymer; for our purposes micelles) tends to migrate to
regions of high stress; if the plateau modulus increases with polymer
concentration ($dG/d\phi>0$), positive feedback enhances concentration
fluctuations perpendicular to the shear compression axis (as seen in
Ref.~\cite{DecLerBer01}) and (sometimes) shifts the spinodal of any
nearby Cahn-Hilliard (CH) demixing instability\cite{ClarMcle98}.

Further evidence for concentration coupling comes from the slight
upward slope~\cite{BRL98} in the stress plateau BF of many wormlike
micellar systems.  This is most readily explained (in planar shear at
least) by a concentration difference between the coexisting
bands~\cite{schmitt95,olmstedlu97}.  Any coupling to concentration has
important implications for the kinetics of macroscopic band formation,
due to the large time scales involved in diffusion.


In this paper we model the initial stages of banding formation in the
unstable regime.  We use the non-local Johnson-Segalman (d-JS)
model~\cite{johnson77,olmsted99a} for the dynamics of the micellar
stress, since this is the simplest tensorial model that admits a flow
curve of negative slope.  To incorporate concentration coupling, we
use a two-fluid model~\cite{brochdgen77,milner91,deGen76,Brochard83}
for the relative motion of the micelles and solvent. We perform a
linear stability analysis (similar in spirit to the CH calculation for
conventional liquid-liquid demixing) for coupled fluctuations in shear
rate, micellar stress and concentration about an initially homogeneous
shear state.  We calculate the ``spinodal'' boundary of the region in
which these fluctuations are unstable. We then consider startup
``quenches'' into the unstable region, predicting the length and time
scales at which inhomogeneity first emerges.  We also discuss the
physical nature of the growing instability, according to whether its
eigenvector is dominated by the flow variables or by concentration.

The two-fluid d-JS model to be studied in this paper was introduced
and analysed briefly by us in a previous letter~\cite{letter}. In this
work we discuss more fully the model's origin and approximations, and
give detailed numerical and analytical arguments supporting the
results announced in Ref.~\cite{letter}.

For simplicity we consider only fluctuation wavevectors in the
velocity gradient direction, $\vect{k}=k\hat{\vect{y}}$, and
(separately) the vorticity direction $\vect{k}=k\hat{\vect{z}}$. (The
stability of the latter in fact turns out to be unaffected by shear in
our model.)  Indeed, the spinodal is commonly {\em defined} using only
these fluctuations for which $\vect{k}.\vect{x}=0$
~\cite{Onuk89}(since any component in the flow gets advected onto the
velocity gradient direction), though this restriction is less relevant
inside the unstable region where fluctuations can grow on a time scale
similar to that of advection.

The paper is structured as follows. In section~\ref{sec:model} we
introduce the model and describe its intrinsic constitutive curves. In
Sec.~\ref{sec:onflowcurve}) we perform a linear stability analysis for
slow shear rate sweeps along the intrinsic constitutive curve, to
define the spinodal onset of instability.  In Sec.~\ref{sec:startup}
we consider shear startup ``quenches'' into the unstable region.  We
conclude in section~\ref{sec:conclusion}.

\section{The Model}
\label{sec:model}

The existing literature contains several approaches for coupling
concentration and
flow~\cite{brochdgen77,HelfFred89,DoiOnuk92,milner93,WPD91,ClarMcle98,beris94}.
The two-fluid model considered by us follows closely that of
Milner~\cite{milner93}, although we extend his work slightly by
including a Newtonian contribution to the micellar stress, for reasons
discussed in Sec.~\ref{sec:coupled_spinodal}.  (Milner was mainly
interested in slow shear phenomena, for which the Newtonian terms are
unimportant.)

The basic assumption of the two-fluid model is a separate force
balance for the micelles (velocity $\vm$) and the solvent (velocity
$\vs$) within any element of solution.  These are added to give the
force balance for the centre of mass velocity
\be
\vect{v}=\phi\vm+(1-\phi)\vs,
\ee
and subtracted for the relative velocity
\be \vrel=\vm-\vs, 
\ee
which in turn specifies the concentration fluctuations.  We give these
dynamical equations in Sec.~\ref{sec:equations} below.  First, we
specify the free energy.

\subsection{Free energy}

In a sheared fluid, one cannot strictly define a free energy because
shear drives the system out of equilibrium. Nonetheless, for realistic
experimental shear rates many of the internal degrees of freedom of a
polymeric solution relax very quickly compared with the rate at which
they are perturbed by the externally moving constraints.  Assuming
that such a separation of time-scales exists, one can effectively
treat these fast variables as equilibrated. By integrating over them,
one can define a free energy for a given fixed configuration of the
slow variables. For our purposes, the relevant slow variables are the
fluid momentum $\rho\vect{v}$ and micellar concentration $\phi$ (which
are both conserved and therefore truly slow in the hydrodynamic
sense), and the micellar strain $\tens{W}$ (which is slow for all
practical purposes). More precisely $\tens{W}$ is the local strain
that would have to be reversed in order to relax the micellar stress:
\be
\tens{W}=\tens{E}.\tens{E}^T-\tens{\delta},\; \mbox{ with }\; \delta\vect{r}'=\tens{E}.\delta\vect{r}
\ee
where $\delta\vect{r}'$ is the deformed vector corresponding to the
undeformed vector $\delta\vect{r}$. 

The resulting free energy is assumed to comprise separate kinetic,
osmotic and elastic components:
\be
\label{eqn:free_energy_total}
F=F^{\rm K}(\vect{v})+F^{\rm o}(\phi)+F^{\rm e}(\tens{W}).
\ee
The kinetic component is
\be
\label{eqn:kinetic_fe}
F^{\rm K}(\vect{v})=\tfrac{1}{2}\int d^3x \rho\vect{v}^2.
\ee
The osmotic component is
  \bea 
  \label{eqn:free_energy}
  F^{\rm o}(\phi)&=&\int d^3x \left[f(\phi)+\frac{g}{2}(\nablu \phi)^2\right]\nonumber\\
         &\approx& \tfrac{1}{2} \int d^3q\, (1+\xi^2 q^2)f''|\phi(q)|^2,
  \eea                             %
  where $f''$ is the osmotic susceptibility and $\xi$ is the
  equilibrium correlation length for concentration fluctuations.
The elastic component is
\be
\label{eqn:elastic_fe}
F^{\rm e}(\tens{W})=\tfrac{1}{2}\int d^3x G(\phi) {\rm tr} \left[\tens{W}-\log(\tens{\delta}+\tens{W})\right]
\ee
in which $G(\phi)$ is the plateau modulus.

\subsection{Dynamical equations}
\label{sec:equations}

We now specify the dynamics. In any fluid element, the forces and
stresses on the {\em micelles} are as follows:
\begin{enumerate}
  
\item The viscoelastic stress $\tens{\sigma}$ of the micellar
  backbone:
%
%
%
\be
\tens{\sigma}=2(\tens{W}+\tens{\delta}).\frac{\delta F^{\rm e}}{\delta \tens{W}}= G(\phi)\tens{W}.
\ee

\item The osmotic force $\phi \nablu [\delta F^{\rm o}/\delta \phi]$
  which induces conventional cooperative micellar diffusion and a
  ``non-linear elastic force'' $\phi \nablu [\delta F^{\rm e}/\delta
  \phi]$.

\item
  
  A Newtonian stress $2\,\phi\, \etam\, \Dm^{0}$ from fast micellar
  relaxations such as Rouse modes, where  
  \be
  \label{eqn:Dm1}
  \Dm^{0}=\Dm-\tfrac{1}{3}\tens{\delta}\,{\rm Tr} \Dm,
  \ee
and
\be
\label{eqn:Dm2}
  \Dm=\tfrac{1}{2}\left[\nablu\, \vm + (\nablu\, \vm)^T\right]. 
\ee
We call $\etam$ the ``Rouse viscosity'' (distinct from the zero shear
viscosity of the {\em total} micellar stress).  $\etam$ is assumed to
be independent of $\phi$, but is prefactored by the extensive factor
$\phi$.

\item
  
  The drag force $\zeta(\phi)\vrel$ impeding the relative motion of
  micelles and solvent. Scaling theory \cite{pgdgpolymer} suggests
  $\zeta\sim 6\pi \eta \xi^{-2}$ where $\eta$ is the mean viscosity
  defined in Eqn.~\ref{eqn:av_visc} below.
  
\item
  
  Stress due to gradients in the hydrostatic pressure $p$.

\end{enumerate}
The overall micellar force balance equation is thus:
\begin{widetext}
\be
\label{eqn:micelle}
\rho_{\rm m}\,\phi\,\left(\partial_t+\vm.\nablu\right)\vm=\nablu.G(\phi)\,\tens{W}-\phi\,\nablu \frac{\delta F(\phi)}{\delta\phi}+2\,\nablu.\,\phi\, \etam\, \Dm^{0} -\zeta(\phi)\,\vrel-\phi\nablu p.
\ee
Likewise, for the solvent we have the Newtonian viscous stress, the
drag force and the hydrostatic pressure:
\be
\label{eqn:solvent}
\rho_{\rm s}(1-\phi)\left(\partial_t+\vs.\nablu\right)\vs=2\nablu.\,(1-\phi)\, \etas\, \Ds^{0}+\zeta(\phi)\vrel-(1-\phi)\nablu p.
\ee
These equations contain the basic assumption of ``dynamical
asymmetry'', \ie\ that the viscoelastic stress acts only on the
micelles and not on the solvent. Adding them, and assuming equal mass
densities $\rho_{\rm m}=\rho_{\rm s}=\rho$, we obtain the overall
force balance equation for the centre of mass motion
\be
\label{eqn:navier}
\rho\,\left(\partial_t+\vv.\nablu\right)\vv -\rho\vrel \vv.\nablu \phi +\rho\phi(1-\phi)\vrel.\nablu\vrel=\nablu.G(\phi)\,\tens{W}-\phi\,\nablu\frac{\delta F(\phi)}{\delta\phi}+2\,\nablu.\,\eta \, \tens{D}^{0} + 2\,\nablu.\,\tilde{\eta} \, \Drel^{0} -\nablu p
\ee
\end{widetext}
in which $\Dv^0$ and $\Drel^0$, defined analogously to $\Dm^0$ in
Eqns.~\ref{eqn:Dm1} and~\ref{eqn:Dm2} above, are the traceless
symmetrised shear rate tensors for the centre of mass and relative
velocity respectively. We have also defined
  \be
  \label{eqn:av_visc}
  \eta=\phi\etam +(1-\phi)\etas 
\ee
and
\be
\tilde{\eta}=\phi(1-\phi)(\etam-\etas).
\ee
The equal and opposite drag forces have cancelled each other in
Eqn.~\ref{eqn:navier}, which is essentially the Navier-Stokes'
equation generalised to include osmotic stresses.  The pressure
$p$ is fixed by incompressibility,
\be
\label{eqn:incomp}
\nablu.\vv=0.
\ee

We attach a cautionary note to Eqn.~\ref{eqn:navier}. The right hand
side (RHS) is the net force acting on the fluid element.  The LHS,
however, equals the usual convective inertial force {\em plus} the two
terms in $\vrel$. We consider this to be an unsatisfactory aspect of
the two-fluid model that is seldom acknowledged in the literature. One
might argue that the separate advected derivatives of
Eqns.~\ref{eqn:micelle} and~\ref{eqn:solvent} should have
$\vect{v}.\nablu\vect{v}_{\rm i}$ in place of $\vect{v}_{\rm
  i}.\nablu\vect{v}_{\rm i}$ (for ${\rm i}\in {\rm m,s}$).  This still
leaves the correction $-\vrel \vv.\nablu\phi$ on the LHS of
Eqn.~\ref{eqn:navier} and does not improve the approximation.  In this
paper, however, we consider only small fluctuations about a
homogeneous shear state (in which $\vrel=0$), and the correction terms
are truly negligible.

Subtracting the micellar and solvent Eqns.~\ref{eqn:micelle}
and~\ref{eqn:solvent} (with each predivided by its own volume
fraction) gives the relative motion, which in turn specifies the
concentration fluctuations:
\begin{widetext}
\be
\label{eqn:concentration}
(\partial_t+\vv.\nablu)\phi=-\nablu.\phi(1-\phi)\vrel=-\nablu.\frac{\phi^2(1-\phi)^2}{\zeta(\phi)}\left[-\nablu\frac{\delta
    F}{\delta \phi}+\frac{1}{\phi}\nablu.
  G(\phi)\tens{W}+2{\mathcal{N}}_1(\Dv^{0})+2{\mathcal{N}}_2(\Drel^{0})\right].
\ee
%
%
${\mathcal{N}}_1$ and ${\mathcal{N}}_2$ are the following
Newtonian terms:
\be
{\mathcal{N}}_1(\Dv^{0})=\frac{\etam}{\phi}\nablu.\phi \Dv^0-\frac{\etas}{1-\phi}\nablu.(1-\phi)\Dv^0
\ee
and
\be
{\mathcal{N}}_2(\Drel^{0})=\left(\frac{\etam}{\phi}+\frac{\etas}{1-\phi}\right)\nablu.\phi(1-\phi)\Drel^{0}.
\ee
We have neglected the inertia in $\vrel$ since it is small compared
with the drag force $\zeta\vrel$~\cite{neglect_inertia}.  The osmotic
contribution to the first term in the square brackets of
Eqn.~\ref{eqn:concentration} shows that micelles diffuse down to
gradients in the chemical potential, as in the usual CH description.
The second term states that micelles diffuse in response to gradients
in the viscoelastic stress. As described by Helfand and
Fredrickon~\cite{HelfFred89}, this provides a mechanism whereby
micelles can diffuse {\em up} their own concentration gradient: the
parts of an extended molecule that are in regions of lower viscosity
will, during the process of relaxing to equilibrium, move more than
those parts mired in a region of higher visosity and concentration.  A
relaxing molecule therefore on average moves towards the higher
concentration region.  This provides a mechanism whereby shear can
enhance concentration fluctuations, and is the essential physics of
the two fluid model.

For the dynamics of the viscoelastic micellar backbone strain we use
the phenomenological d-JS model~\cite{johnson77,olmsted99a}: 
%
\begin{equation}
\label{eqn:JSd}
(\partial_t+\vm.\nablu)\tens{W}=a(\Dm.\tens{W}+\tens{W}.\Dm)+(\tens{W}.\Omm-\Omm.\tens{W})+2\Dm-\frac{\tens{W}}{\tau(\phi)}+\frac{l^2}{\tau(\phi)} \nablu^2 \tens{W},
\end{equation}
\end{widetext}
where $2\Omm=\nablu \vm - (\nablu \vm)^T$ with $(\nablu
\vm)_{\alpha\beta}\equiv \partial_{\alpha}(v_{\rm m})_\beta$.
$\tau(\phi)$ is the Maxwell time. The length $l$ could, for example,
be set by the mesh-size or by the equilibrium correlation length for
concentration fluctuations. Here we assume the former, since the
dynamics of the micellar conformation are more likely to depend on
gradients in molecular conformation than in concentration. The
equilibrium correlation length $\xi$ of course still enters our
analysis, through the concentration free energy of
Eqn.~\ref{eqn:free_energy}.  Together, $l$ and $\xi$ set the length
scale of any interfaces.
In the absence of concentration coupling this non-local term involving
$l$ is needed to reproduce a steady banded state with a uniquely
selected stress~\cite{olmsted99a}, although other
treatments~\cite{Yuan99b,Dhon99} have used alternative forms for
non-local terms that also give a uniquely selected stress. The slip
parameter $a$ measures the non-affinity of the molecular deformation,
\ie\ the fractional stretch of the polymeric material with respect to
that of the flow field. For $|a|<1$ (slip) the intrinsic constitutive
curve in planar shear is capable of the non-monotonicity of
Fig.~\ref{fig:schem}.  We use
Eqns.~\ref{eqn:navier},~\ref{eqn:incomp},~\ref{eqn:concentration}
and~\ref{eqn:JSd} as our model for the remainder of the paper.

\subsection{Flow geometry. Boundary conditions}

We consider idealised planar shear bounded by infinite plates at
$y=\{0,L\}$ with $(\vect{v},\nablu v, \nablu \wedge \vect{v})$ in the
$(\hat{\vect{x}},\hat{\vect{y}},\hat{\vect{z}})$ directions.  The
boundary conditions at the plates are as follows.  For
the velocity we assume there is no slip. For the concentration we assume
\be
\label{eqn:BCconc}
\partial_y \phi=\partial^3_y \phi=0.
\ee
which ensures (in zero shear at least) zero flux of concentration at
the boundaries.  
%
%
Following~\cite{olmsted99a}, for the micellar strain we assume  
\be
\label{eqn:BCstrain}
\partial_y W_{\alpha\beta}=0\;\forall\; \alpha,\beta.
\ee
Conditions~\ref{eqn:BCconc} and~\ref{eqn:BCstrain} together ensure
zero concentration flux at the boundary even in shear.  For controlled
shear rate conditions (assumed throughout)
\be
\label{eqn:constant_strain_rate}
\bar{\gdot}=\int_0^L dy \gdot(y)={\rm constant.}
\ee

\subsection{Model parameters}
\label{sec:parameters}

\begin{table}
  \begin{center}
    \begin{tabular}{|p{2.85cm}|c|c|c|}
     \hline
     {\small\bf Parameter} & {\small \bf Symbol} $Q$ & {\small \bf Value at $\phi=0.11$} & $\frac{d\log Q}{d\log\phi}$ \\
     \hline\hline
     {\small Rheometer gap} & $L$ & $0.15\mbox{\,mm}$  & 0 \\
     \hline
     {\small Maxwell time} & $\tau$ & $0.17\mbox{\,s}$ & 1.1 \\
     \hline
     {\small Plateau modulus} & $G$ & $232\mbox{\,Pa}$ & 2.2 \\
     \hline
     {\small Density} & $\rho$ & $10^3\mbox{\,kg\,m}^{-3}$ & 0 \\
     \hline
     {\small Solvent viscosity} & $\etas$ & $10^{-3}\mbox{\,kg\,m}^{-1}\mbox{s}^{-1}$ & 0 \\
     \hline
     {\small Rouse viscosity} & $\etam$ & $0.4\mbox{\,kg\,m}^{-1}\mbox{s}^{-1}$ & 0 \\
     \hline
     {\small Mesh size} & $l$ & 2.6$\times 10^{-8}\mbox{m}$ & -0.73 \\
     \hline
     {\small Diffusion coefficient} & $D$ & $3.5 \times 10^{-11}\mbox{m}^2\mbox{s}^{-1}$ &0.77 \\
     \hline
     {\small Drag coefficient} & $\zeta$ & $2.4\times10^{12}\mbox{kg\,m}^{-3}\mbox{\,s}^{-1}$ & 1.54 \\
     \hline
     {\small Correlation length} & $\xi$ & $6.0\times10^{-7}$m & -0.77 \\
     \hline
     {\small Slip parameter} & $a$ & $0.92$ & 0 \\
     \hline
    \end{tabular}
  \end{center}
 \caption{Experimental values of the model's parameters at volume fraction $\phi=0.11$ (column 3). Scaling laws for the dependence of each parameter upon $\phi$ (column 4). In most calculations we use  the reference values of column 3 at $\phi=0.11$, then tune $\phi$ using the scaling laws of column 4. Only where  stated do we allow the parameters to vary independently.  \label{table:parameters}}
\end{table}

Values for the model parameters are taken as follows. We assume the
solvent viscosity $\etas$ and density $\rho$ to be those of water. We
take the plateau modulus $G$ and the Maxwell time $\tau$ from linear
rheology~\cite{lerouge_note} at $\phi=0.11$ on CTAB(0.3M)/${\rm
NaNO}_3\rm {(1.79M)}$/H$_2$O. We estimate the Rouse viscosity $\etam$
from the (limited data on the) high shear branch of the flow curve of
a closely related system~\cite{lerouge_note}. The mesh size is
estimated to be $l\approx (k_BT/G)^{1/3}$~\cite{pgdgpolymer}. In fact
this form is only truly valid for a good solvent, although in the
interests of simplicity we assume it to be good approximation even for
systems closer to demixing. We take the diffusion coefficient $D$ and
the equilibrium correlation length $\xi$ from dynamic light scattering
(DLS) data~\cite{CanHirZan85} on CTAB/KBr/H$_2$O, at a comparable
micellar volume fraction. We calculate the drag coefficient
$\zeta=6\pi\eta \xi^{-2}$~\cite{pgdgpolymer}.  We fix the slip
parameter $a=0.92$ by comparing our intrinsic constitutive curve in
the semi-dilute regime to that of Cates' model for wormlike
micelles~\cite{cates90}.  We then have realistic values for all
parameters, at $\phi=0.11$ (table~\ref{table:parameters}).

Exploring this large parameter space is a daunting prospect so we
shall not, in general, vary the parameters independently. Instead we
simply tune the single parameter $\phi$, relying on known semi-dilute
scaling laws for the dependence of the other parameters upon $\phi$
(column 4 of table~\ref{table:parameters}).  For simplicity we assume
that the slip parameter $a$ is independent of $\phi$. We rescale
stress, time and length so that $G(\phi=0.11)=1$, $\tau(\phi=0.11)=1$,
and $L=1$, where $L$ is the rheometer gap ($0.15{\rm mm}$) used in
Ref.~\cite{lerouge_note}. We also often eliminate $\rho$ in favour of
the Reynolds time $\tau_{\rm d}=\rho L^2/\eta_s$.  In total the model
has $8$ scaled parameters.

\subsection{intrinsic constitutive curves}
\label{sec:flow_curves}

\begin{figure}[h]
\centering
\includegraphics[scale=0.3]{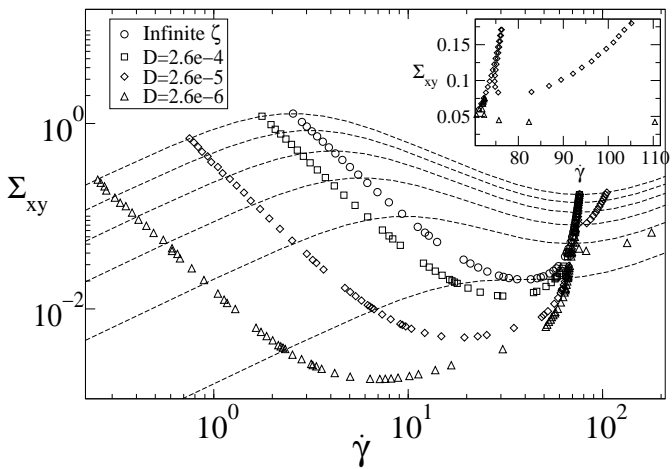}
\caption{intrinsic constitutive curves for $\phi=0.11$,$0.091,0.072,0.053,0.034,0.015$ (dashed lines, downwards). Spinodals for: the uncoupled limit $\zeta\to\infty$ ($\circ$); coupled model with
  $D(\phi=0.11)$ taken from the DLS data
  (table~\ref{table:parameters}) ($\square$); coupled model with $D$
  artificially reduced ($\lozenge$,$\vartriangle$).  Inset: enlargement at
  high $\gdot$.
\label{fig:spinodals_with_phi} } 
\end{figure}

In planar shear, the stationary homogeneous solutions to
Eqns.~\ref{eqn:navier},~\ref{eqn:incomp},~\ref{eqn:concentration}
and~\ref{eqn:JSd} for given $\gdot$ and $\phi$ are $\vrel=0$ and
\bea
\label{eqn:intrinsic_flow_curve}
W_{xy}&=&\frac{\gdot\tau(\phi)}{1+b\gdot^2\tau^2(\phi)},\nonumber\\
W_{yy}&=&-\frac{1}{(1+a)}\frac{b\gdot^2}{1+b\gdot^2}\nonumber\\
W_{xx}&=&\frac{1+a}{a-1}W_{yy},\nonumber\\
W_{zz}&=&W_{xz}=W_{yz}=0,
\eea
where
\be
b=1-a^2.
\ee
The total shear stress is the sum of the micellar stress and a
Newtonian component:
\be
\label{eqn:intrinsic_flow_curve1}
\Sigma(\gdot,\phi)=G(\phi)W_{xy}+\etab(\phi)\gdot.
\ee
This defines a set of intrinsic constitutive curves
$\Sigma(\gdot,\phi)$ (dashed lines in
Fig.~\ref{fig:spinodals_with_phi}). The criterion for the
non-monotonicity of $W_{xy}$ to dominate the Newtonian term
$\etab(\phi)\gdot$ and cause non-monotonicity in the overall stress
$\Sigma$ is
%
$\etab(\phi)<\tfrac{1}{8}G(\phi)\tau(\phi)$.
%
As $\phi$ is reduced, therefore, the region of negative slope narrows,
terminating in a ``critical'' point at $\phi=\phi_{\rm c}\approx
0.015$.  The same qualitative trend has been seen in
CPCl/NaSal/brine~\cite{berret94b}.

\section{Uncoupled limit; Instabilities; Positive feedback}
\label{sec:uncoupled_limit}

In the limit of infinite drag, \ie\ $\zeta\to\infty$ at fixed $f''(\phi)$,
the relative motion between micelles and solvent is switched off,
disabling concentration fluctuations. In the slightly different limit
of $\zeta\to\infty$ at fixed micellar diffusion coefficient
\be
D=\frac{\phi^2(1-\phi)^2 f''}{\zeta},
\ee
the concentration still fluctuates, but independently of the rheological
variables. Eqn.~\ref{eqn:concentration} then reduces to the CH
equation (with a $\phi$-dependent mobility).
Independently of $\delta\phi$, the shear rate and micellar stress
together obey uniform-$\phi$ d-JS
dynamics~\cite{radulescu99a,olmsted99a,radulescu99b}.
Accordingly, two separate instabilities are possible:

\begin{enumerate}
\item {\underline{Demixing instability}}
  
  For $D<0$, concentration has its own CH demixing instability,
  governed primarily by the free energy defined in
  Eqn.~\ref{eqn:free_energy}. In this work we consider only {\em flow
    induced} instabilities, for which $D>0$.

\item {\underline{Mechanical instability}}
  
  For shear rates where the intrinsic constitutive curve has negative
  slope $d\Sigma/d\gdot<0$, fluctuations in shear rate and micellar
  stress have their own shear banding instability, which for
  convenience we call ``mechanical''. However, we emphasize that the
  term ``mechanical instability'' is often misused in the literature.
  For example, the origin of banding in semi-dilute micelles is
  usually described as purely ``mechanical''.  In contrast banding in
  concentrated systems, nearer an equilibrium isotropic-nematic (I-N)
  transition, is generally described as a flow induced perturbation to
  this I-N transition. However, such a transition possesses an essentially identical ``mechanical'' instability so
  there is no sharp distinction between these scenarios.  Our approach
  is more obviously relevant to semi-dilute systems since we rely
  purely on non-linear coupling between shear and $\tens{W}$ to induce
  instability, rather than on a perturbation of a non-linear free
  energy $F^{\rm e}(\tens{W})$.  Accordingly, {\em but for convenient
    nomenclature only}, we refer to this instability as
  ``mechanical''.

\end{enumerate}

For finite drag, these instabilities mix. Consider, for example, the
second term in the square brackets of Eqn.~\ref{eqn:concentration},
which encourages micelles to move up gradients in the micellar
backbone stress.  If the micellar stress increases with concentration
($dG/d\phi>0$), positive feedback occurs leading to shear enhancement
of concentration
fluctuations~\cite{brochdgen77,HelfFred89,DoiOnuk92,milner93,WPD91} or
(from the opposite extreme) concentration-coupled enhancement of a
flow instability. Instability can then occur even if $D>0$ and
$d\Sigma/d\gdot>0$, so the domain of (what was the purely) mechanical
instability is broadened relative to that in which $d\Sigma/d\gdot<0$.
For systems far from a zero-shear CH demixing instability (\ie\ for
$D\gg 0$) the mixed instability is ``mainly mechanical'' (dominated by
$\delta\gdot,\delta\tens{W}$). For systems that in zero shear are
already close to demixing ($D\gae 0$) the instability sets in at low
shear rate, where it is essentially demixing triggered by flow
(dominated by $\delta\phi$).  Of course, as in the above discussion of
mechanical \versus I-N instabilities, there is no sharp distinction
between these two extremes: the model captures a smooth crossover
between them.
  
Enhancement of flow instabilities by concentration coupling was first
predicted by the remarkable insight of Schmitt {\it et
al.}~\cite{schmitt95}.  The feedback mechanism described in the
previous paragraph for our model corresponds to the direct assumption
by Schmitt {\it et al.}~\cite{schmitt95} of a chemical potential
$\mu=\mu(\gdot)$.  However this is only truly equivalent to our
approach if the viscoelastic stress $\tens{W}(\gdot)$ can adjust
adiabatically (assumed in~\cite{schmitt95}). We find below that the
dynamics inside the spinodal {\em are} dictated by the rate of
micellar stress response.  The spinodal is unaffected, since response
here is adiabatic by definition. Schmitt {\it et al.} also predicted
an instability for negative feedback, but concluded it to be similar
in character to a pure mechanical instability in which concentration
plays no role.  In our model, negative feedback would correspond to
$dG/d\phi<0$; here we consider only positive feedback (in the language
of Ref.~\cite{schmitt95}, $C>0$).

In this work, we model the onset of instability for two different flow
histories. The first (section~\ref{sec:onflowcurve}) assumes an
initial state on the intrinsic constitutive curve, and is used to
define the ``spinodal'' limit of stability in sweeps along this flow
curve.  This is analogous to defining the spinodal of a van der Waals
fluid via quasistatic compression, and suffers the same practical
ambiguity that {\em finite} fluctuations can cause separation/banding
via {\em metastable} kinetics before the spinodal is reached.  The
second history (section~\ref{sec:startup}) considers a startup
``quench'' into the unstable region, and is (essentially) the
counterpart of a temperature quench into the demixing regime of a van
der Waals fluid.  The analysis here is complicated by the fact that
the fluctuations emerge against the time-dependent startup flow.

\section{Initial condition on intrinsic constitutive curve; slow shear rate sweeps}
\label{sec:onflowcurve}

\subsection{Linear analysis}

We encode the system's state as follows,
\be
\label{eqn:pseudovec}
\vect{u}=\gdot\hat{\vect{e}}_{\gdot}+\sum_{ij}W_{ij}\hat{\vect{e}}_{W_{ij}}+\phi\,\hat{\vect{e}}_\phi,
\ee
in which the $\hat{\vect{e}}$ are
dimensionless unit vectors~\cite{vector_caution}.
%
%
Consider fluctuations about a mean initial state $\bar{\vect{u}}$ that
is on the intrinsic constitutive curve:
\be
\label{eqn:linearize}
\vect{u}(\vect{r},t)=\bar{\vect{u}}+\sum_{\vect{k}} \delta \vect{u}_{\vect{k}}(t) \exp\left(i\vect{k}.\vect{r}\right)
\ee
where $\bar{\vect{u}}=[\bar{\gdot},\bar{\tens{W}},\bar{\phi}]$ with
$\bar{\tens{W}}(\bar{\gdot})$ the stationary homogeneous solution
given by Eqn.~\ref{eqn:intrinsic_flow_curve}. The sum in
Eqn.~\ref{eqn:linearize} covers positive and negative $\vect{k}$, with
$\vect{u}_{-\vect{k}}=\vect{u}^*_{\vect{k}}$ ensuring
$\vect{u}(\vect{r},t)$ is real. Linearising
Eqns.~\ref{eqn:navier},~\ref{eqn:incomp},~\ref{eqn:concentration}
and~\ref{eqn:JSd} in these fluctuations, we find
\be
\label{eqn:langevin}
\partial_t\delta \vect{u}_{\vect{k}}(t)=\tens{M}_{\vect{k}}.\delta\vect{u}_{\vect{k}}(t).
\ee
%
%
The stability matrix $\tens{M}_{\vect{k}}$ depends on the model
parameters and the initial state
$\bar{\vect{u}}=[\bar{\gdot},\bar{\tens{W}},\bar{\phi}]$. Its eigenmodes are determined by
\be
\label{eqn:find_modes}
\omega_{\vect{k},\alpha}\eigenvec=\tens{M}_{\vect{k}}\eigenvec
\ee
where $\alpha$ is the mode index. The functions
$\omega_{\vect{k},\alpha}$ versus $\vect{k}$ define a multi-branched
dispersion relation. For the initial state $\bar{\vect{u}}$ to be
stable, all dispersion branches must be negative. A positive
eigenvalue $\omega_{\vect{k},\alpha}$ indicates an unstable mode that
grows exponentially in time with relative order-parameter amplitudes
specified by the corresponding eigenvector $\eigenvec$. In an upward
sweep along the intrinsic constitutive curve, the lower spinodal lies
where the eigenvalue $\omega_{\vect{k}^*}$ with the largest real part
(maximised over $\vect{k}$ and $\alpha$) crosses the imaginary axis in
the positive direction.  The upper spinodal is defined likewise, for
sweeps towards the unstable region from above.  Strictly, only
harmonics of the gap-size are allowed; but in order to define the
spinodal independently of rheometer geometry we allow arbitrarily
small wavevectors. For any shear rate between the spinodals, the
dispersion relation is positive for some range of wavevectors.
Typically, we find just one unstable dispersion branch
$\omega_{\vect{k}}$ (although we comment below on an exception, for
some model parameters, at very high shear rates).

Below we give numerical results for this unstable branch, focusing on
any global maximum, which would indicate a selected length scale
$\vect{k}^{*-1}$ at which inhomogeneity emerges most quickly. We also
give results for the unstable eigenvector
${\tt{\vect{v}}}_{\vect{k}^*}$ at this maximum.  As noted above, we
consider only fluctuations of wavevectors $\vect{k}=k
\,\hat{\vect{y}}$ and (separately) $\vect{k}=k \,\hat{\vect{z}}$. In
fact the stability of fluctuations $\vect{k}=k \,\hat{\vect{z}}$ turns
out to be unaffected by shear in our model (see
section~\ref{sec:vorticity}) so we study in detail only $\vect{k}=k
\,\hat{\vect{y}}$.

For $\vect{k}=k \,\hat{\vect{y}}$, $\delta v_y=0$ by
incompressibility. Fluctuations in the remaining variables decouple
into three independent subspaces:
\begin{itemize}
\item
  
  ${\mathfrak{S}_1}\equiv [ik\delta v_x=\delta\gdot,\delta W_{xy},\delta
  W_{xx}, \delta W_{yy}, \delta \phi]$

\item

${\mathfrak{S}_2}=[ik\delta v_z,W_{xz},W_{yz}]$ 

\item

 ${\mathfrak{S}_3}=W_{zz}$.

\end{itemize}
In all unstable regimes, for this flow history, only
${\mathfrak{S}_1}$ is unstable.  (In {\em startup} at high shear
rates, ${\mathfrak{S}_2}$ can go unstable; however it is always less
unstable than ${\mathfrak{S}_{1}}$ in the relevant time window; see
section~\ref{sec:startup}.)  Accordingly, we focus on
${\mathfrak{S}_1}$. For convenience, we change variables to
\be 
\label{eqn:defineZ}
Z=\frac{a-1}{2}W_{xx}+\frac{1+a}{2}W_{yy} 
\ee
and
\be
\label{eqn:defineY}
Y=\frac{a-1}{2}W_{xx}-\frac{1+a}{2}W_{yy}.
\ee
In sections~\ref{sec:uncoupled_fc} and~\ref{sec:coupled_fc} we give
our numerical results for the instability in this subspace
${\mathfrak{S}_1}$. In some regimes we also give {\em qualitative}
analytical results, obtained using the following simplified stability
matrix of ${\mathfrak{S}_1}$,
\bw
\be 
\label{eqn:fluct_in_y_matrix_conc}
 \tens{M}_{\vect{k}}=    \left( 
    \begin{array}{cc}
      \begin{array}{ccc} 
\D -\frac{\eta k^2}{\eta_s \tau_d} & \D -\frac{k^2}{\eta_s\tau_d} & 0 \\[10truept]
     \D   1+\bar{Z} &\D -1-l^2k^2 & \bar{\gdot} \\[10truept]
    \D    -b\bar{W}_{xy} &\D -b\bar{\gdot} &\D -1-l^2k^2 \\[10truept]\hline
     0        & 0 & \D k^2/\tilde{\zeta}  
     \end{array}
    & 
 \left|
    \begin{array}{c} 
      \D -\frac{G'\bar{W}_{xy}k^2}{\eta_s\tau_d} \\[10truept]
      \D\bar{W}_{xy}\tau'  \\[10truept]
      \D \bar{Z}\tau'  \\[10truept] \hline
    \D -\tilde{D}k^2(1+\xi ^2k^2)
    \end{array}
  \right.
    \end{array}
  \right)
\vspace{-0.25cm}
\begin{array}{c} {\scriptstyle \delta\gdot}\\[10truept] {\scriptstyle \delta W_{xy}} \\[10truept]
    {\scriptstyle \delta Z}  \\[10truept] {\scriptstyle \delta \phi} 
\end{array}
\vspace{0.75cm}
\ee
\ew
with
\be
\tilde{\zeta}=\frac{(1+a)}{\phi(1-\phi)^2}\zeta
\ee
and
\be
\tilde{D}=D-\frac{\bar{Z}G'}{\tilde{\zeta}}.
\ee
%
%
($\bar{Z}$ is negative so $\tilde{D}>D$.)
Matrix~\ref{eqn:fluct_in_y_matrix_conc} is exact in the uncoupled
limit $\zeta\to\infty$. For finite $\zeta$ it contains several
approximations~\cite{neglect_Y} (most notably neglecting $\delta Y$)
and so underestimates the growth rate of the coupled instability;
however the qualitative trends are unaffected.  In some places below
we further neglect terms of order $\eta$. This is only valid for
concentrations not to near the critical concentration $\phi_{\rm c}$
and shear rates not too far above the lower spinodal, so that
$\etab\gdot \ll GW_{xy}$. In any case, startup at higher shear rates
is too violent to study experimentally~\cite{lerougeprivate}.

\subsection{Results: uncoupled limit}
\label{sec:uncoupled_fc}

In the limit $\zeta\to\infty$ at fixed $f''/\zeta$, fluctuations in
the rheological variables decouple from those in concentration and the
stability matrix is exactly
\be 
\label{eqn:fluct_in_y_matrix_no_conc}
\tens{M}_{\vect{k}}= 
  \left( \begin{array}{cc}
      \tens{M}_{\rm M} & --\\
         0      & -Dk^2(1+\xi ^2k^2) \end{array} \right)
 \hspace{0.5cm}
\vspace{-0.25cm}
\begin{array}{c} {\scriptstyle {\rm mechanical}} \\[10truept] {\scriptstyle \delta \phi} 
\end{array}
\vspace{0.75cm}
\ee
in which $\tens{M}_{\rm M}$ is the upper-left $3\times 3$
``mechanical'' sector of matrix~\ref{eqn:fluct_in_y_matrix_conc}.  The
three elements represented by the dash in
matrix~\ref{eqn:fluct_in_y_matrix_no_conc} are non-zero, but
irrelevant since all off-diagonal elements in the bottom row are zero.
In this section, we give results for the spinodal and dispersion
relation in this uncoupled limit.

\subsubsection{Spinodal}
\label{sec:uncoupled_spinodal}

%

For each of a range of concentrations, we calculated spinodals
numerically: see the circles in figures~\ref{fig:spinodals_with_phi}.
The unstable region coincides with that of negative constitutive slope
$d\Sigma/d\gdot$, as expected, and vanishes at a ``critical point''
$\phi_{\rm_c}\approx 0.015$, as in the experiments of
Ref.~\cite{berret94b}.

Analytically, the eigenvalues of the stability
matrix~\ref{eqn:fluct_in_y_matrix_no_conc} obey the quartic equation:
\be
\label{eqn:polynomial}
\omega_k^4 + a\omega_k^3 + b \omega_k^2 + c\omega_k + d=0
\ee
where $d=\mbox{Det }\tens{M}_{\vect{k}}$. The roots of any polynomial
with real coefficients are either real, or complex-conjugate pairs.
This gives two possibilities for the spinodal.  First, the root with
the largest real part could be zero, implying the onset of a
monotonically growing instability; this corresponds to the 3-sub-space
(of the 4-space spanned by $a,b,c,d$) for which $d=0,a>0,b>0,c>0$.
Alternatively, the root could be one of a pure imaginary pair,
implying the onset of growing oscillations; this also corresponds to a
3-sub-space though not (in general) defined simply by one of the
$a,b,c,d$ axes.  For the parameters considered, we have mostly found
the first case.\footnote{In the {\em coupled} model a rather
  pathological oscillatory instability occurs at very high shear rates
  (Sec.~\ref{sec:coupled_spinodal}).}  Accordingly, our analysis
hereafter considers only this first case for which the spinodal is
given by $d=\mbox{Det }\tens{M}_{\vect{k}}=0$. (In all parameter
regimes studied, this automatically ensures $a>0,b>0,c>0$.)  It can
further be shown that $\mbox{Det }\tens{M}_{\vect{k}}<0$ in the
unstable region, \ie\ 
\be
\label{eqn:uncoupled_spinodal0}
D{\mathfrak{D}}_{\rm M}>0
\ee
in which
\bea 
\label{eqn:JSdet}
{\mathfrak{D}}_{\rm M} &\equiv& \mbox{Det }\tens{M}_M\nonumber\\
&=& \frac{k^2}{\eta_s \tau_d} \left\{-\eta(1+b\bar{\gdot} ^2)-(1+\bar{Z})+b\bar{W}_{xy}\bar{\gdot}\right\}\nonumber\\
&=& -\left[\frac{k^2}{\eta_s
    \tau_d}(1+b\bar{\gdot}^2)\right]\frac{d\bar\Sigma}{d\bar{\gdot}}.
\hspace{0.5cm} \eea
To a good approximation we have neglected the interfacial terms in
calculating the spinodal: they merely cut off the dispersion relation
at short length scales without affecting the sign of the maximum
growth rate.  The term in the square bracket of Eqn.~\ref{eqn:JSdet}
is always positive, so the condition for instability is finally just
\be
\label{eqn:uncoupled_spinodal}
-D\frac{d\bar\Sigma}{d\bar{\gdot}}>0.
\ee
For an increasing flow curve $\frac{d\bar\Sigma}{d\bar{\gdot}}>0$, CH
$\phi-$demixing can occur for $D<0$. As noted above, in this paper we
consider only $D>0$, for which the unstable region is
$\frac{d\bar\Sigma}{d\bar{\gdot}}<0$, as shown numerically.  Here, the
instability occurs in the upper $3\times 3$ subspace of the
matrix~\ref{eqn:fluct_in_y_matrix_no_conc}, and is purely mechanical.
Although the normal stresses (encoded by $Z=Z(W_{xx},W_{yy})$) have
apparently cancelled from Eqn.~\ref{eqn:uncoupled_spinodal}, they in
fact play a crucial role, as follows.  The origin of the instability
is the term $\tfrac{k^2}{\etas\tau_d}b\bar{W}_{xy}\bar{\gdot}$ in the
curly braces of Eqn.~\ref{eqn:JSdet}. In this term,
$\tfrac{-k^2}{\etas\tau_d}$ is the prefactor to $\delta W_{xy}$ in the
$\delta\gdot$ equation, and states that a local increase in $W_{xy}$
causes a (diffusive) decrease in $\gdot$. The remaining factor feeds
back positively, by stating that a decrease in $\gdot$ in turn tends
to increase $W_{xy}$, consistent with the negative slope in the
constitutive curve. However this factor itself comprises two
subfactors, each of which describes a mechanism that explicitly
involves the normal stress, $Z$. The first, $-b\bar{W}_{xy}$, states
that the decrease in $\gdot$ causes an increase in $Z$. The second,
$\bar{\gdot}$ states that this increase in $Z$ causes an increase in
$\gdot$, thereby completing the positive feedback.  This role of
normal stress was not considered in early studies of mechanical
instability~\cite{Yerushalmi70}. Note finally that the {\em absolute}
values of the normal stresses are important, not just the difference
$W_{yy}-W_{xx}$: the trace of the {\em micellar contribution} to the
stress tensor is not arbitrary.

\subsubsection{Dispersion relation}
\label{sec:uncoupled_dispersion}

\begin{figure*}[tbp]
  \centering \subfigure[$\zeta\to\infty$.]{
    \includegraphics[scale=0.35]{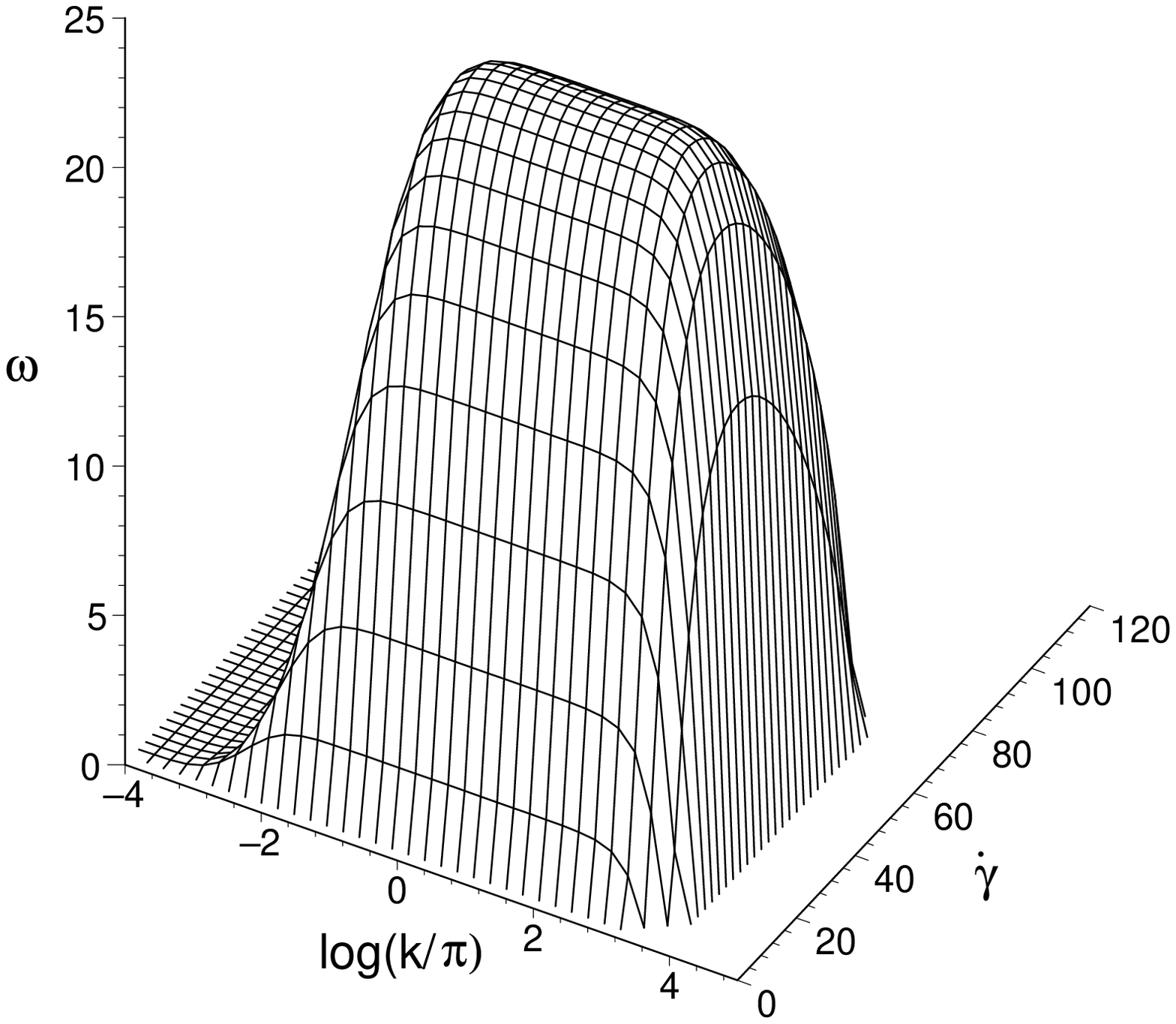}
\label{fig:dispersion_phi0.11:a}
}
\subfigure[$D(\phi=0.11)=2.6\times 10^{-4}$.]{
\includegraphics[scale=0.35]{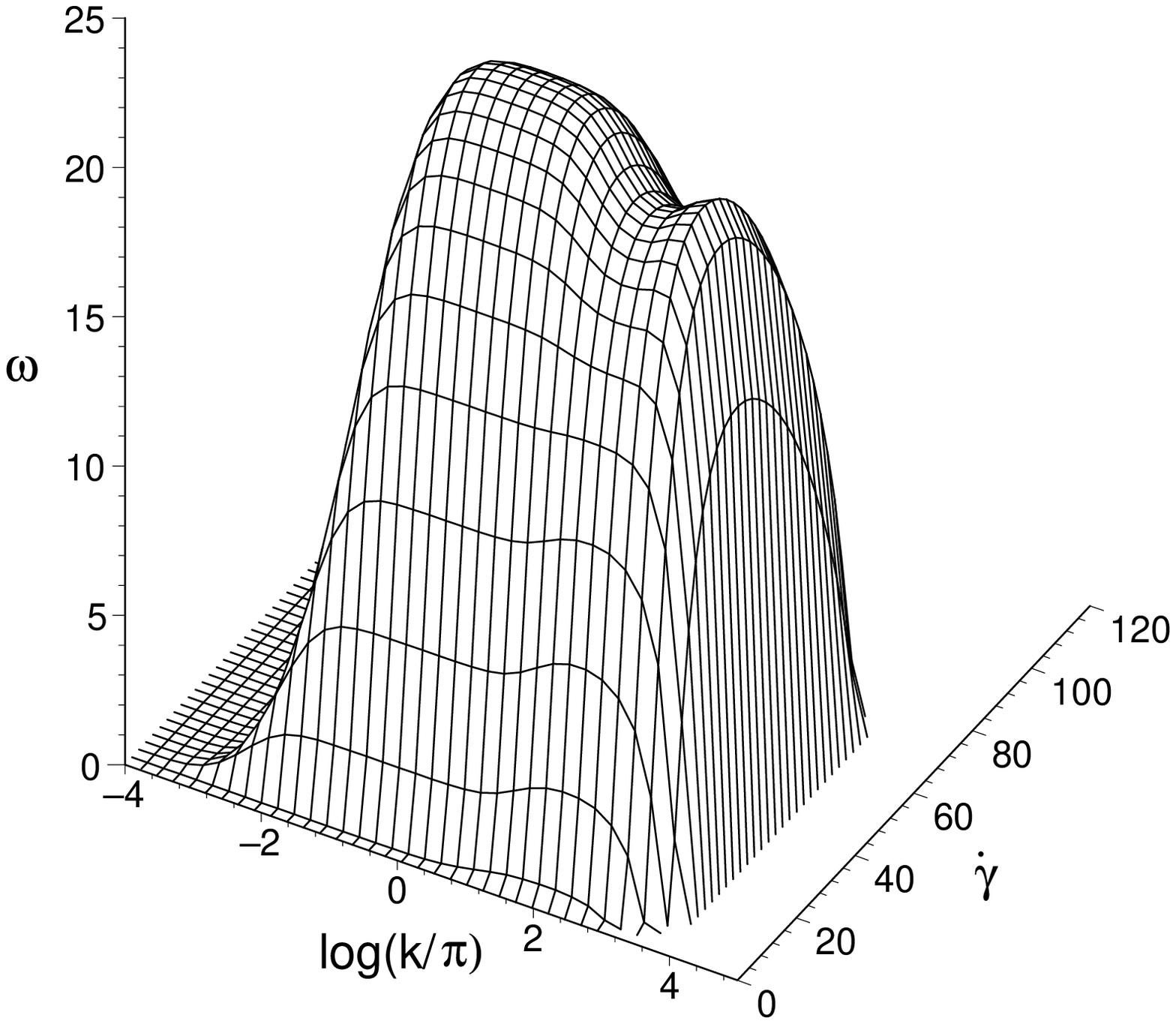}
\label{fig:dispersion_phi0.11:c}
}
\subfigure[$D(\phi=0.11)=2.6\times 10^{-6}$.]{
\includegraphics[scale=0.35]{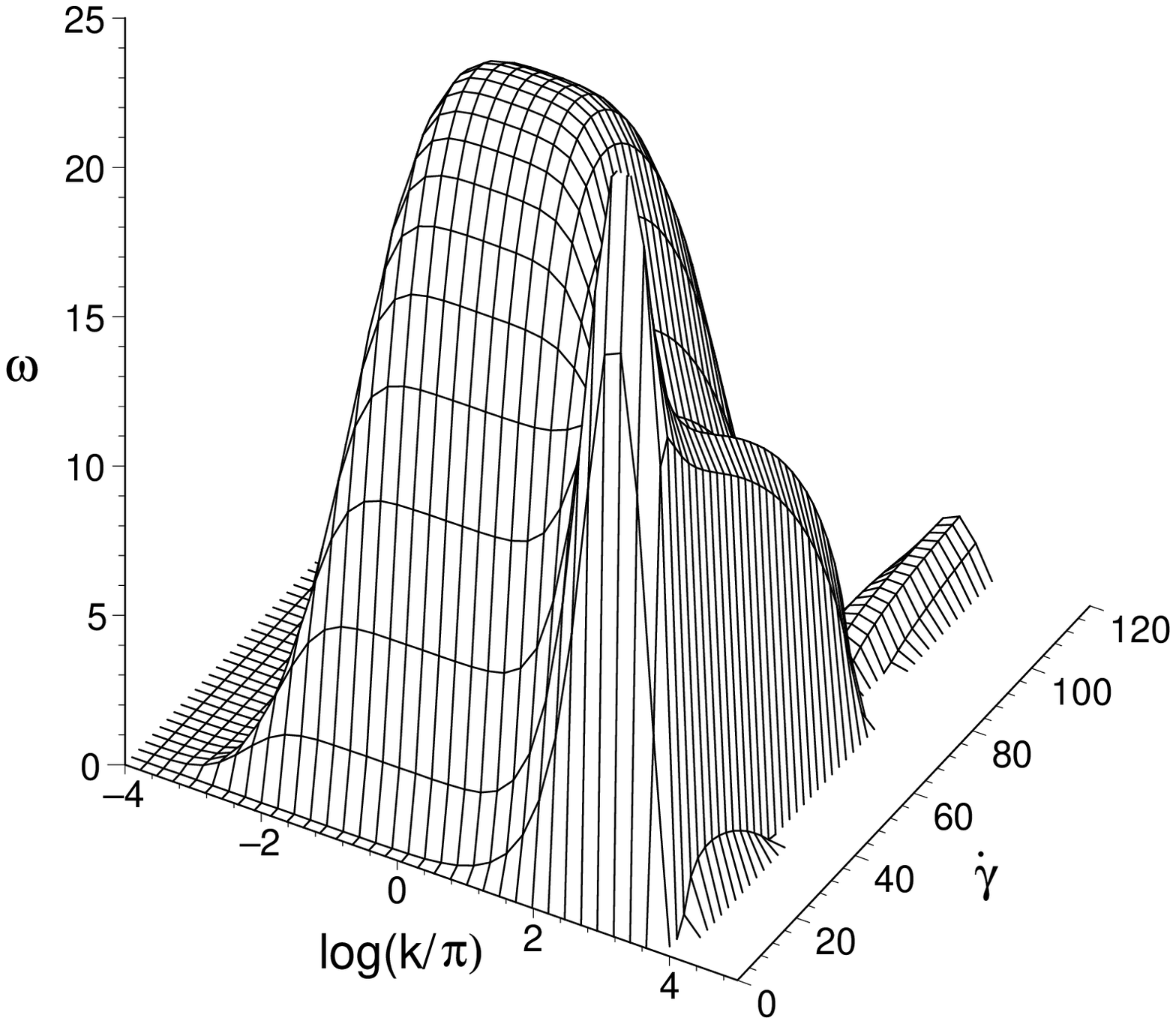}
\label{fig:dispersion_phi0.11:e}
}
\subfigure[$\zeta\to\infty$ (detail).]{
  \includegraphics[scale=0.35]{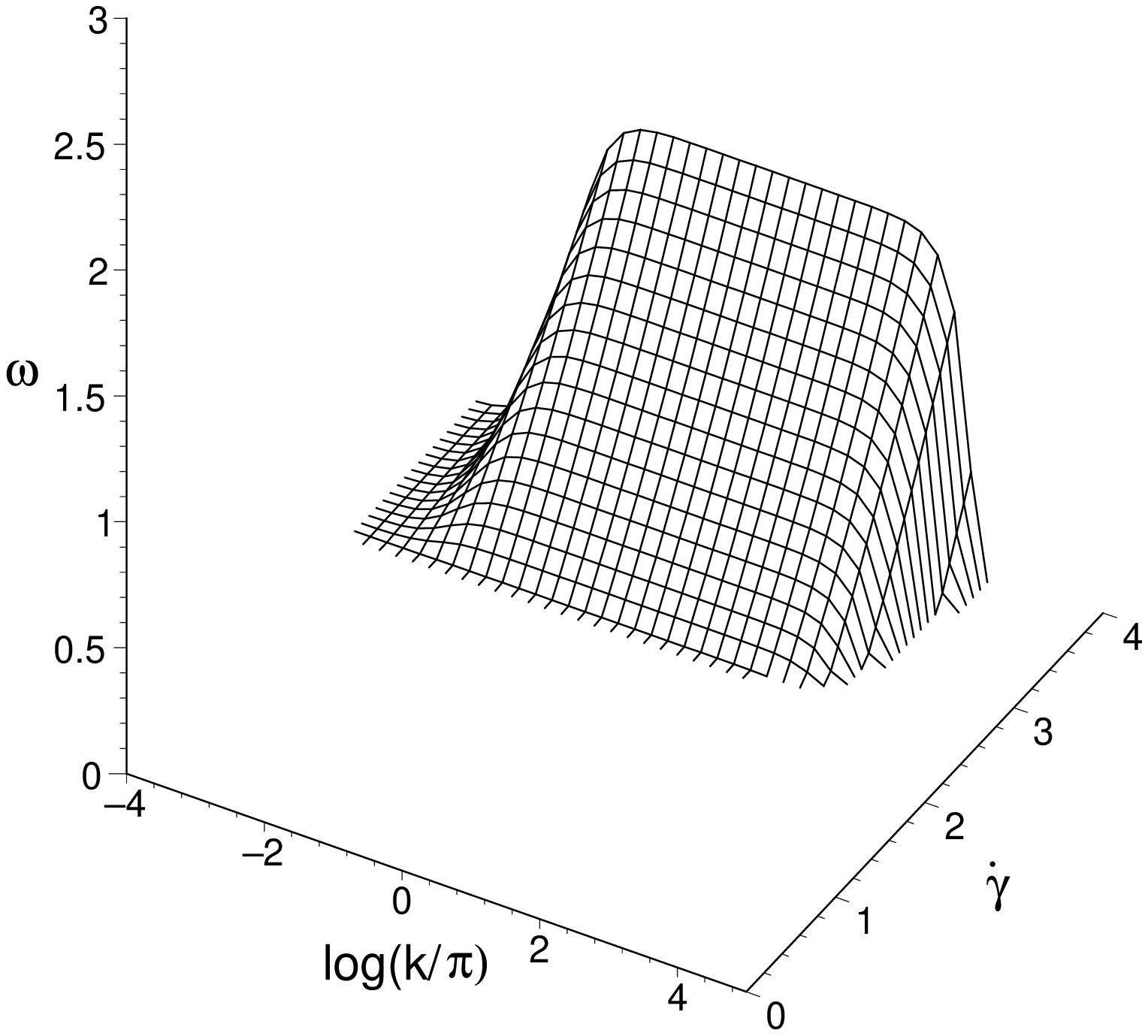}
\label{fig:dispersion_phi0.11:b}
}
\subfigure[$D(\phi=0.11)=2.6\times 10^{-4}$ (detail).]{
  \includegraphics[scale=0.35]{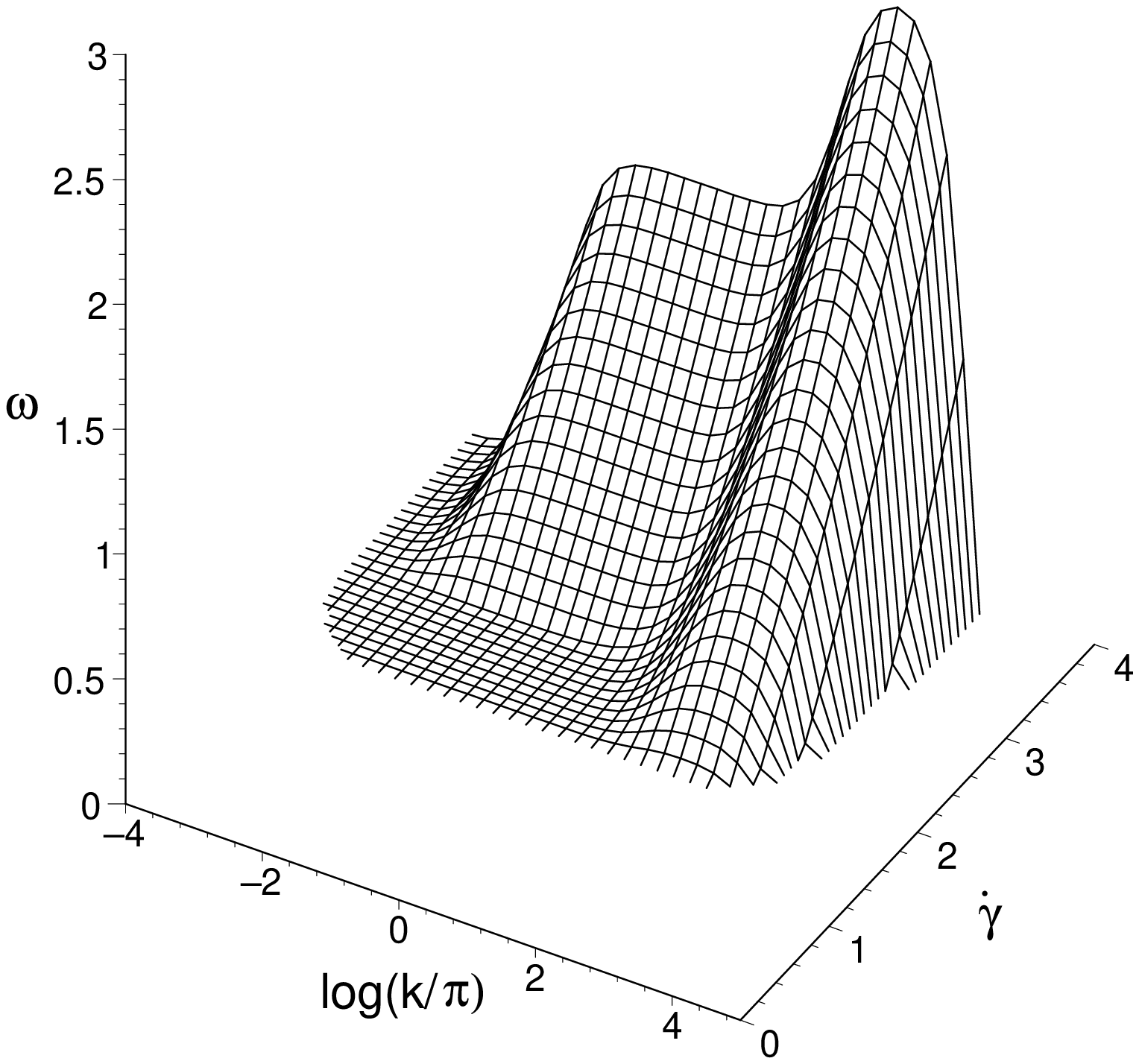}
\label{fig:dispersion_phi0.11:d}
}
\subfigure[$D(\phi=0.11)=2.6\times 10^{-6}$ (detail).]{
  \includegraphics[scale=0.35]{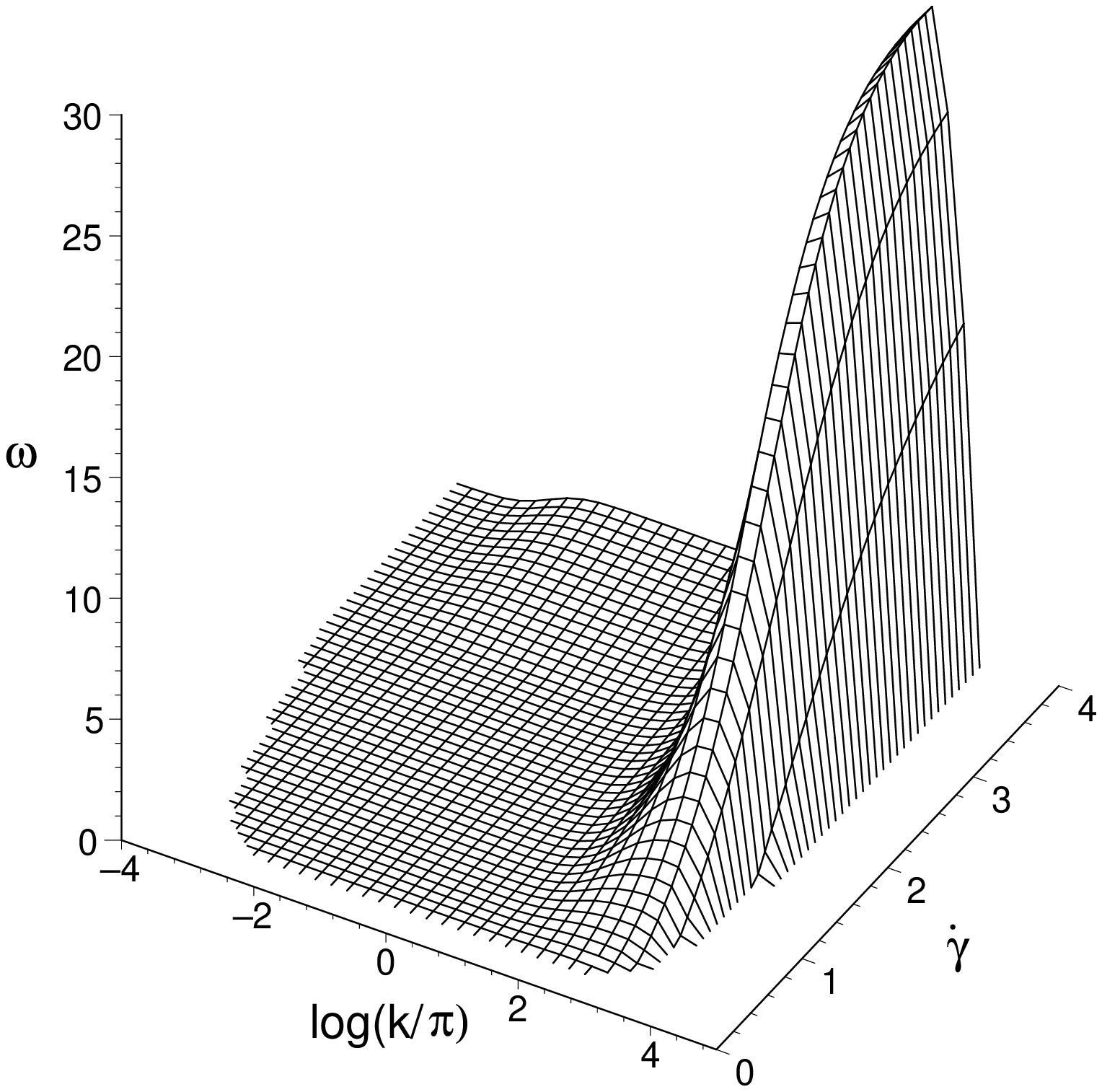}
\label{fig:dispersion_phi0.11:f}
}
\caption{Positive (unstable) dispersion branch at $\phi=0.11$.~\ref{fig:dispersion_phi0.11:a} and \ref{fig:dispersion_phi0.11:b} are for the uncoupled model; \ref{fig:dispersion_phi0.11:c} and \ref{fig:dispersion_phi0.11:d} are for the coupled model in which all parameters assume the experimental values of table~\ref{table:parameters} (spinodal given by $\square$s in Fig.~\ref{fig:spinodals_with_phi}); \ref{fig:dispersion_phi0.11:e} and \ref{fig:dispersion_phi0.11:f} are for a coupled model in which $D(\phi)$ is artificially  reduced  (spinodal  given by $\vartriangle$s in Fig.~\ref{fig:spinodals_with_phi}). For each vertical pair of graphs, the bottom is an enlargement of the top one, at shear rates near the lower spinodal. In each subfigure, the white space defines $(\gdot, k)$ values for which all  dispersion branches are negative.}
\label{fig:dispersion_phi0.11}. 
\end{figure*}

\begin{figure*}[htbp]
  \centering \subfigure[$\zeta\to\infty$.]{
\label{fig:dispersion_phi0.02:a}
  \includegraphics[scale=0.35]{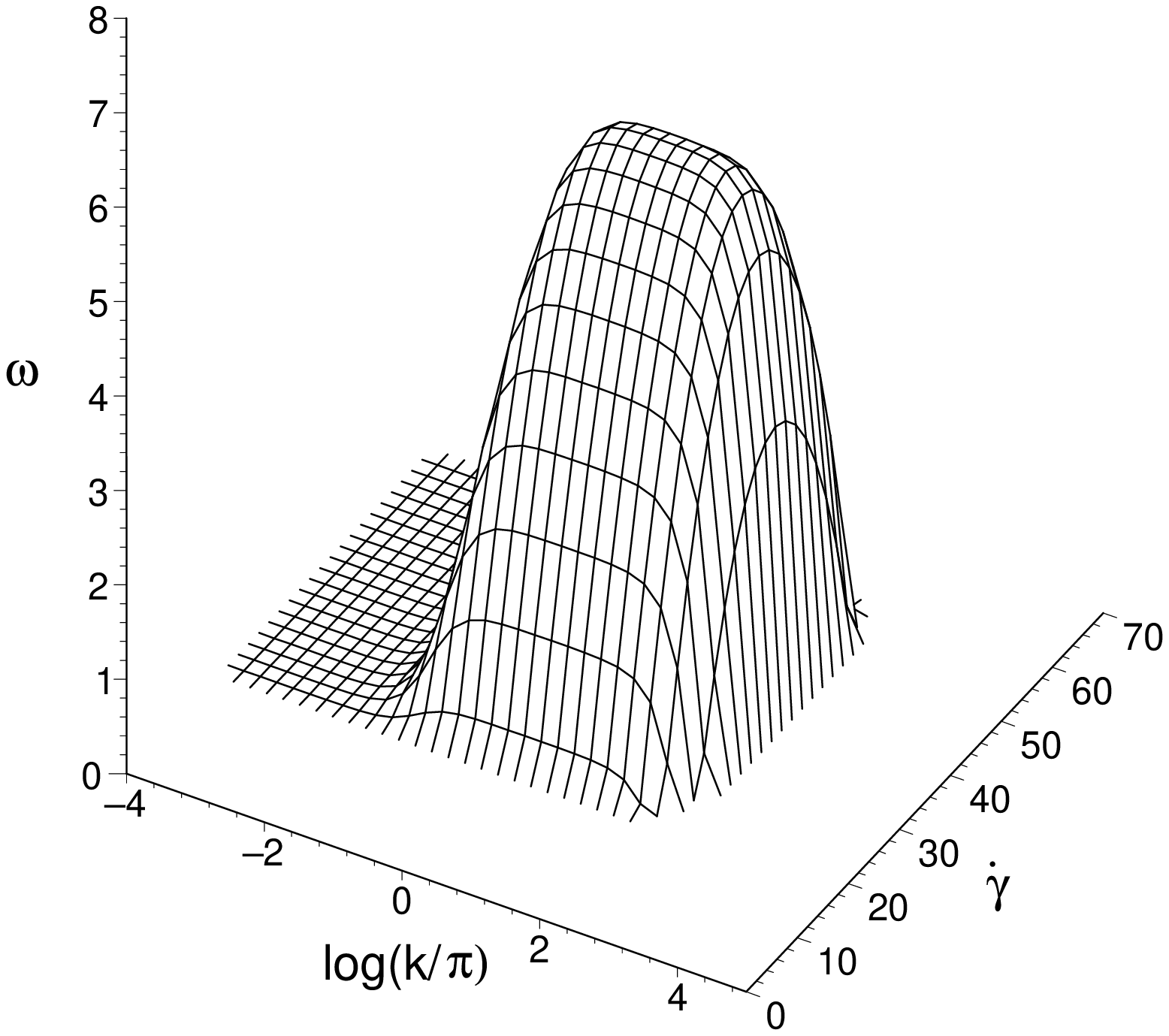}
}
\subfigure[$D(\phi=0.11)=2.6\times 10^{-4}$.]{
\label{fig:dispersion_phi0.02:b}
  \includegraphics[scale=0.35]{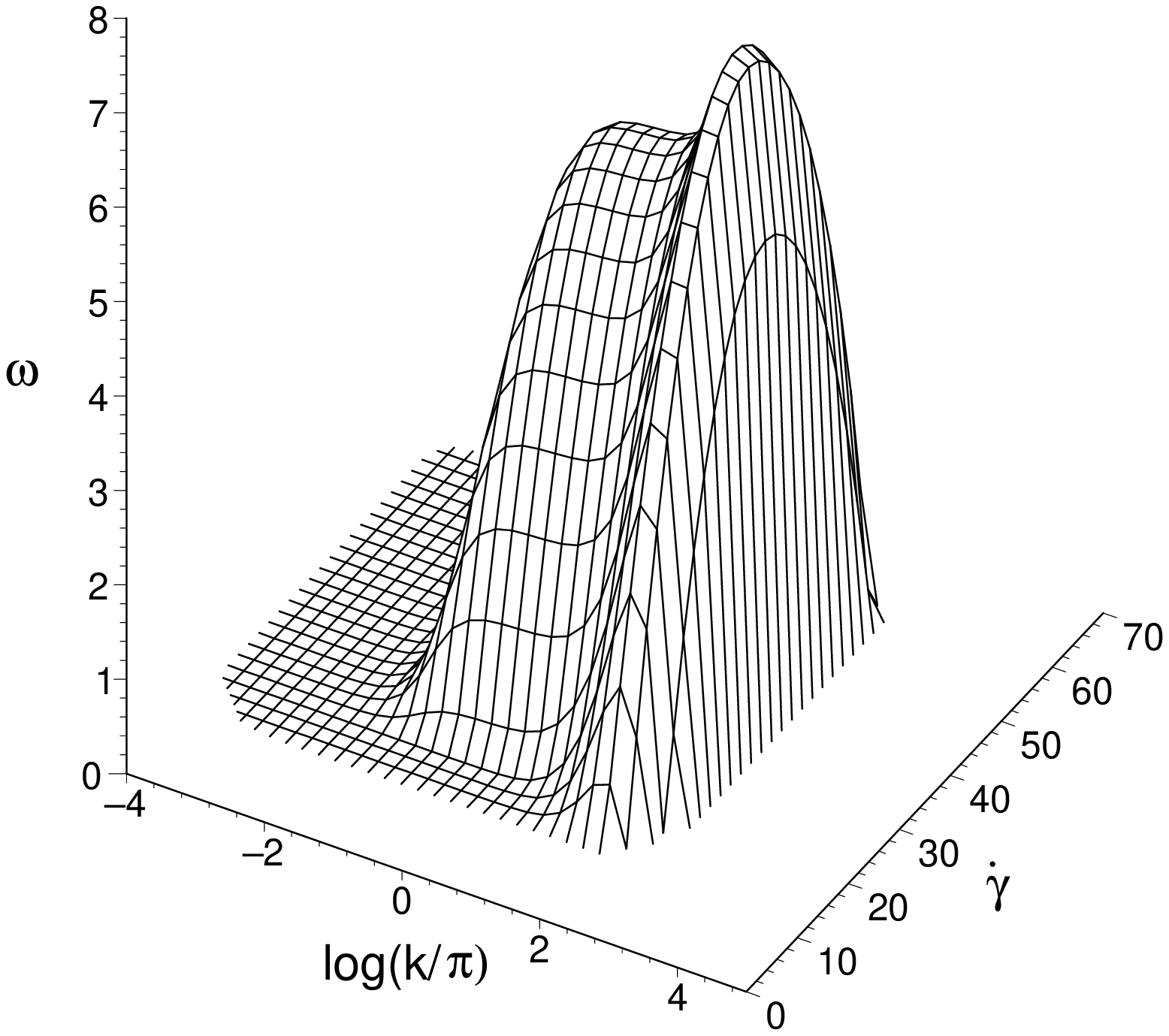}
}
\subfigure[$D(\phi=0.11)=2.6\times 10^{-6}$.]{
\label{fig:dispersion_phi0.02:c}
  \includegraphics[scale=0.35]{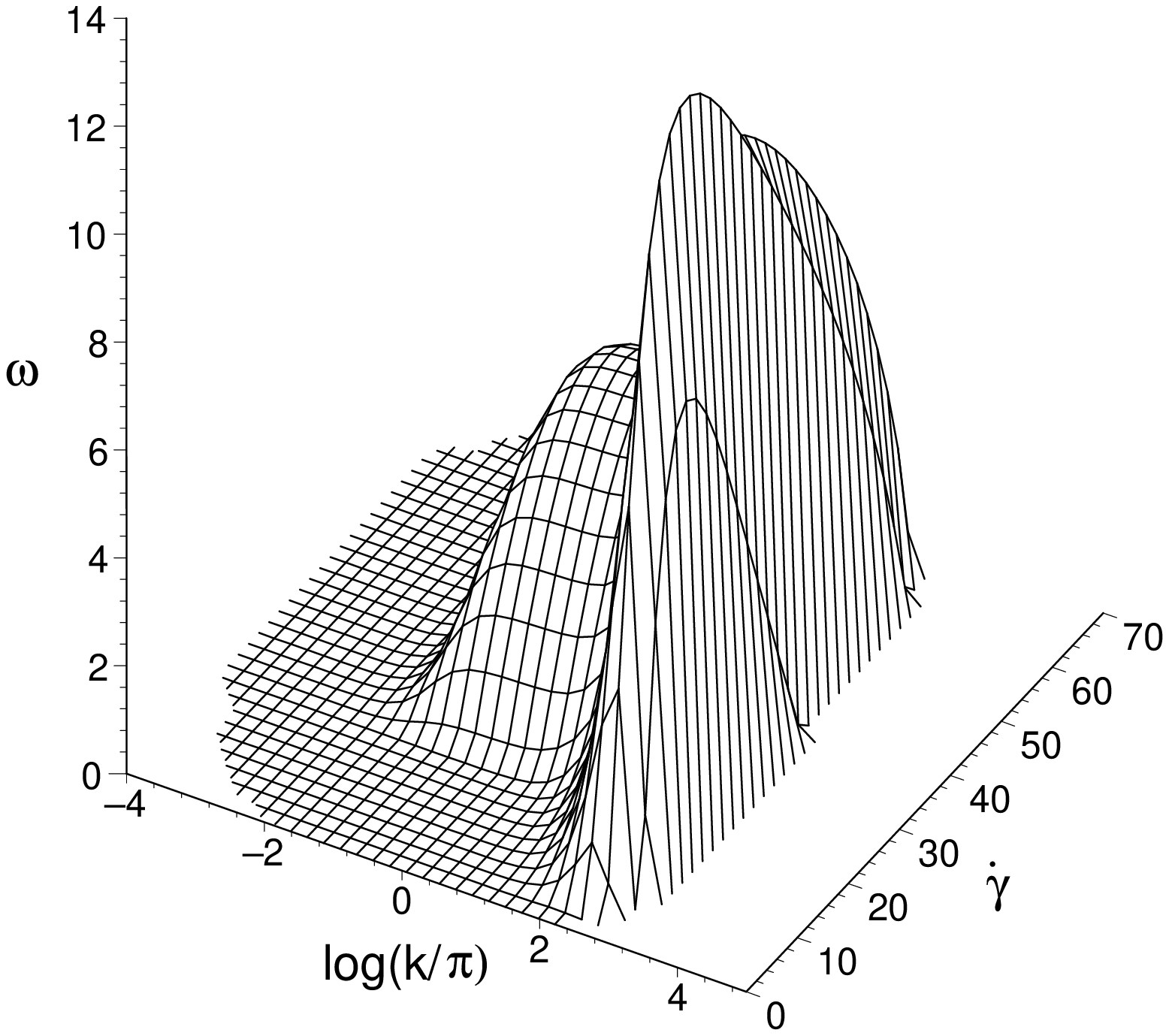}
}
\caption{Unstable dispersion branch at $\phi=0.02$. \ref{fig:dispersion_phi0.02:a} is for the uncoupled model; \ref{fig:dispersion_phi0.02:b} is for the coupled model in which at $\phi=0.11$ all parameters assume the experimental values of table~\ref{table:parameters} (spinodal is given by $\square$s in Fig.~\ref{fig:spinodals_with_phi}); \ref{fig:dispersion_phi0.02:c} is for a coupled model in which $D(\phi=0.11)$ is artificially reduced (spinodal given by $\vartriangle$s in Fig.~\ref{fig:spinodals_with_phi}). In each subfigure, the white space defines $(\gdot, k)$ values for which all dispersion branches are negative.}
\label{fig:dispersion_phi0.02}
\end{figure*}


\begin{figure*}[htbp]
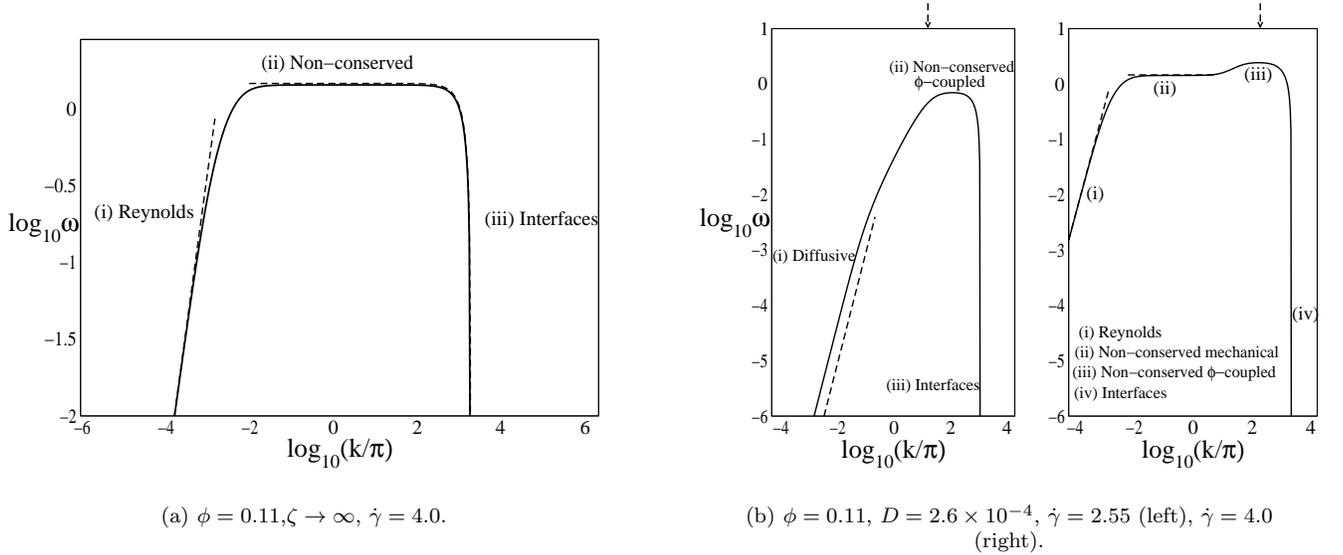

  \centering \subfigure[$\phi=0.11$,$\zeta\to\infty$, $\gdot=4.0$.]{
\label{fig:dispersion_raw_phi0.11_gdot4.0_2D}
\includegraphics[scale=0.29]{dispersion_raw_phi0.11_gdot4.0_2D_asymp.eps}
}\hspace{1.0cm}
\subfigure[$\phi=0.11$, $D=2.6\times 10^{-4}$, $\gdot=2.55$ (left), $\gdot=4.0$ (right).]{
\label{fig:dispersion_coupled_phi0.11_gdot2.0_and_4.0_2D}
  \includegraphics[scale=0.29]{dispersion_coupled_phi0.11_gdot2.0_and_4.0_2D_asymp.eps}
}
\caption{Illustration of the various dispersion regimes discussed in the text. (a) for the uncoupled mechanical instability (b) for the coupled model. (b,left) is for a shear rate that  would be stable in the uncoupled limit $\zeta\to\infty$; (b,right) is for a higher shear rate that {\em is} inside the uncoupled mechanical spinodal. The dashed lines are the approximate analytical asymptotes given in Eqns.~\ref{eqn:Reynolds_asymp},~\ref{eqn:JSplatsimple} and~\ref{eqn:dispersion_coupled_regime1}. The dashed arrows show the approximate $k^*$ of Eqn.~\ref{eqn:dispersion_peak}. 
\label{fig:analytical_dispersion}
}
\end{figure*}

Before discussing the dispersion relation we make the following
cautionary remark.  This dispersion relation describes fluctuations
about a homogeneous state on the intrinsic constitutive curve.  While we
correctly employed it to define the spinodal {\em boundary} of
instability, it is less useful {\em inside} the unstable region since
one cannot prepare an unstable initial state.  Indeed, startup
quenches into the unstable region in general go unstable long before
the intrinsic constitutive curve can be attained (see
section~\ref{sec:startup}).  However, the main features of this
dispersion relation do still appear in their time dependent
counterparts of startup flow.  Our motivation for discussing them here
is to gain early qualitative insight without the complication of
time-dependence.

For this uncoupled model (with this initial condition) we only observe
{\rm one} positive dispersion branch, shown in
Figs.~\ref{fig:dispersion_phi0.11:a} for $\phi=0.11$ and
\ref{fig:dispersion_phi0.02:a} for $\phi=0.02$. Strictly, only
harmonics $k=n\pi$ of the gap size $L\equiv 1$ are allowed.  However
we show $k<\pi$ as well, because for some systems the features of this
domain (discussed below) could lie in the allowed region $k\ge
\pi$. Fig.  \ref{fig:dispersion_phi0.11:b} contains the same data as
~\ref{fig:dispersion_phi0.11:a}, but enlarged on shear rates near the
lower spinodal: this is the only regime in which startup kinetics have
been studied experimentally since they become too violent at higher
shear rates~\cite{lerougeprivate}.

For a given unstable applied shear rate $\bar\gdot$, the growth rate
$\omega$ tends to zero as $k\to 0$ and as $k\to \infty$, with a broad
plateau in between.  This can be understood via the following
analytical results obtained from the characteristic equation of
matrix~\ref{eqn:fluct_in_y_matrix_no_conc}, and schematised in
Fig.~\ref{fig:dispersion_raw_phi0.11_gdot4.0_2D}.

\begin{description}
  
\item[(i)]
  
  \underline{Reynolds regime $k\to 0$.} Here we find
\be
\label{eqn:Reynolds_asymp}
\omega_k= -\frac{d\bar{\Sigma}}{d\bar{\gdot}}\frac{k^2}{\etas\tau_d}.
\ee
This is marked as a dashed line in
Fig.~\ref{fig:dispersion_raw_phi0.11_gdot4.0_2D}, and agrees well with
the numerical data. Here, the instability is limited by the Reynolds
rate at which shear rate (conserved overall) diffuses a distance
$O(1/k)$: the micellar stress responds adiabatically in comparison.

\item[(ii)]
  
  \underline{Non-conserved plateau regime.} At these shorter length
  scales (but still with $k^2l^2\ll 1$) the growth rate is instead
  limited by the Maxwell time on which the micellar backbone stress
  evolves (the Reynolds number is then effectively zero). Because
  micellar stress is non conserved, the growth rate is independent of
  $k$:
\be
\label{eqn:JSplatsimple}
\omega=\frac{\tdjs}{1+\bar{Z}}= -\frac{1}{(1+\bar{Z})^2}\frac{d\bar{\Sigma}}{d\bar{\gdot}}+O(\eta,\tilde{\eta})
\ee
with
\be
\tdjs=\djs \frac{\eta_s\tau_d}{k^2},
\ee
which is marked as a dashed line (also incorporating the interfacial
regime, below) in Fig.~\ref{fig:dispersion_raw_phi0.11_gdot4.0_2D}.

\item[(iii)]
  
  \underline{Interfacial cutoff.} The dispersion relation is cut off
  once $kl=O(1)$ by the reluctance to form interfaces.  Here,
  $w$ follows from~\ref{eqn:JSplatsimple} with
  $\omega\to\omega+l^2k^2$.

\end{description}

The crossover between the first two regimes occurs at a length scale
much greater than the interfacial cutoff, giving a broad intermediate
plateau.  The maximum in $\omega(k)$ is very shallow and its length
scale exceeds the system size for the experimental systems considered
here.  Therefore fluctuations grow equally quickly at all length
scales from the system size down to the interface width, and there is
no selected length scale.  In a previous work~\cite{Dhon99} that
considered the mechanical instability of a simple model (with no
concentration coupling), a wavevector was apparently selected.
However, there the viscoelastic stress was assumed to respond
adiabatically so the intermediate plateau was absent.

\subsection{Results: coupled model}
\label{sec:coupled_fc}

For finite drag, fluctuations in the mechanical variables are coupled
to those in concentration $\delta \phi$ via two main mechanisms.  The
first (already discussed briefly) involves the 2nd term in the square
brackets of Eqn.~\ref{eqn:concentration}, which decrees that
concentration diffuses in response to gradients in $W_{yy}$ at rate
$\propto 1/\zeta$; the elastic part of the stress
(Eqn.~\ref{eqn:navier}) then increases in proportion to
$G'(\phi)\equiv dG(\phi)/d\phi$, giving positive feedback $\propto
G'/\zeta$. The second (neglected in our analytical work, as already
noted~\cite{neglect_Y}) comes from the elastic contribution $\nablu
(\delta F^{\rm e}/\delta \phi)$ to the {\em first} term in the square
brackets of Eqn.~\ref{eqn:concentration}.

\subsubsection{Spinodal}
\label{sec:coupled_spinodal}

The mechanical instability is enhanced by this concentration coupling.
For the model parameter values of table~\ref{table:parameters} the
spinodals are shifted only slightly (squares in
Fig.~\ref{fig:spinodals_with_phi}).  However the shift increases near
an underlying CH demixing instability, as illustrated by reducing
$D(\phi=0.11)$ at fixed coupling $G'/\zeta$ (diamonds and triangles in
Fig.~\ref{fig:spinodals_with_phi}).  This is intuitively obvious: when
$D$ finally goes negative (not shown) demixing must occur even in zero
shear. In the opposite extreme $D\to\infty$, the uncoupled limit is
recovered (circles). The shifts in the lower spinodal have important
implications for fast upward stress sweep experiments, since ``top''
jumping should in this case occur  {\em before} the maximum of the
underlying flow curve is reached.

An approximate analytical condition for instability that qualitatively
reproduces the shifts in the lower spinodal (found by setting
$\omega_k=0$ in the characteristic equation of the approximate
stability matrix~\ref{eqn:fluct_in_y_matrix_conc}) is
\be
\label{eqn:spinodal_js_phi}
\tilde{D}{\mathfrak{D}}_{\rm M}+\frac{1}{\tilde{\zeta}} {\mathfrak{D}}_{\rm F}>0
\ee
in which ${\mathfrak{D}}_{\rm M}$ is the mechanical determinant
already defined in Eqn.~\ref{eqn:JSdet} and
${\mathfrak{D}}_{\rm F}$ is a ``feedback determinant'',
\bea
\label{eqn:feedback_deter}
{\mathfrak{D}}_{\rm F}&=&\left| 
                       \begin{array}{ccc} 
                             \D 0 & \D -\frac{k^2}{\eta_s\tau_d} & \D -\frac{G'\bar{W}_{xy}k^2}{\eta_s\tau_d} \\[10truept]
                             \D 1+\bar{Z}        & \D -1 & \D \bar{W}_{xy}\tau' \\[10truept]
                             \D -b\bar{W}_{xy}   & \D -b\bar{\gdot} & \D \bar{Z} \tau'
                       \end{array}
                   \right| +O(\eta,\tilde{\eta})\nonumber\\
                            &= &-\frac{k^2}{\eta_s\tau_s}G'\bar{W}_{xy}\left\{-b\gdot(1+\bar{Z})-b\bar{W}_{xy}\right\} +O(\eta,\tilde{\eta})\nonumber\\
                            &= &-G'\bar{W}_{xy}\left[\frac{k^2}{\eta_s\tau_s}(1+b\gdot^2)\right]\frac{d\bar{Z}}{d\bar{\gdot}} +O(\eta,\tilde{\eta}),
\eea
where $d\bar{Z}/d\bar{\gdot}<0$. (The terms in $\tau'$ cancel each
other.) As for the uncoupled model, the interfacial terms have been
neglected in locating the spinodal.  Our final condition for
instability is thus
\be
\label{eqn:final_coupled_spinodal}
\tilde{D}\frac{d\bar\Sigma}{d\bar{\gdot}}+\frac{G'\bar{W}_{xy}}{\tilde{\zeta}}\frac{d\bar{Z}}{d\bar{\gdot}}<0,
\ee
which reduces to the uncoupled condition~\ref{eqn:uncoupled_spinodal}
for $\zeta\to\infty$ at fixed $D$, as required.  The size of the
second term above (which encodes feedback) relative to the
``diagonal'' product of uncoupled instabilities (first term) is set by
$G'/(D\zeta)\sim G'/f''$, \ie\ the ratio of the ``feedback
elasticity'' $G'$ to the osmotic elasticity $f''$.  The kinetic
coefficient $\zeta$ has cancelled from this ratio, since the
instability occurs adiabatically at the spinodal.
Eqn~\ref{eqn:final_coupled_spinodal} corresponds to Eqn. 24 in the
paper of Schmitt {\it et al.}~\cite{schmitt95}.

On the basis of these results, we classify systems into two basic
types.

\begin{itemize}
  
\item \underline{Type I systems} are far from a CH demixing
  instability ($D\gg 0$). The mechanical spinodal is
  shifted only slightly by concentration coupling. 
  
\item \underline{Type II systems} are close to a CH
  instability ($D\gae 0$). The mechanical spinodal is
  strongly perturbed by concentration coupling. 

\end{itemize}
Correspondingly, we anticipate two types of instability (with a smooth
crossover in between):
\begin{itemize}
  
\item \underline{Type A instabilities}, which are essentially
  mechanical (eigenvector mostly in $\delta\gdot$, $\delta\tens{W}$)
  but perturbed by coupling to $\delta\phi$.  These are expected in
  all type I systems; and in type II systems for shear rates well
  above the lower spinodal.
  
\item \underline{Type B instabilities}, which are essentially CH in
  character (eigenvector dominated by $\delta\phi$), but perturbed by
  a coupling to flow.  These occur in type II systems at shear rates
  just inside the lower spinodal: see
  Refs.~\cite{brochdgen77,HelfFred89,DoiOnuk92,milner93,WPD91,ClarMcle98}.
  
\end{itemize}
This intuition is confirmed by the results given in
section~\ref{sec:startup} below. 

The results of Fig.~\ref{fig:spinodals_with_phi} also reveal a second
lobe of instability that appears at high shear rates for small values
of $D$.  However its existence and location are highly sensitive to
the choice of model parameters and to the precise details of model
definition: it appears much more readily and extends to much higher
shear rates if the Newtonian contribution to the micellar stress is
not included.  Its eigenvector is overwhelmingly dominated by $\delta
\gdot$. It is associated with two {\em complex} eigenvalues with equal
positive real parts. We do not study this instability in detail, but
return in section~\ref{sec:conclusion} to discuss its potential
implications.

\begin{figure}[h]
\centerline{\psfig{figure=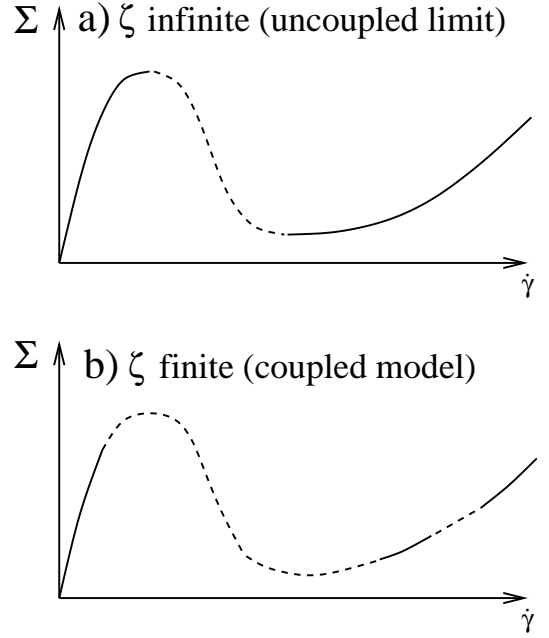,width=7cm}}
\caption{Sketch of the unstable (dashed) region of a mechanical instability a) decoupled  from or b) coupled to concentration. As discussed in the text, concentration coupling broadens the region of instability, and can sometimes cause a new region of instability to develop in the high shear rate branch.
\label{fig:cartoon} } 
\end{figure}

The effect of concentration coupling in broadening the region of
instability is schematized in Fig.~\ref{fig:cartoon}, along with the
possibility of a further region of instability in the high shear rate
branch,  discussed in the previous paragraph.

\subsubsection{Dispersion relation}
\label{sec:coupled_dispersion}

We now discuss the dispersion relation of the coupled model (though
the cautionary remark made in Sec.~\ref{sec:uncoupled_dispersion}
above for the uncoupled model still applies).  We focus mainly, and
firstly, on shear rates above, but quite close to, the lower spinodal,
since this is the regime that is studied experimentally. Comparing the
dispersion relation for the pure mechanical instability
(Fig.~\ref{fig:dispersion_phi0.11:a}) to that for a coupled model of
type I (Fig.~\ref{fig:dispersion_phi0.11:d}), we see that {\em
  concentration coupling enhances the mechanical instability at short
  wavelengths, thereby selecting a length scale}.  We discuss this
length scale in more detail below. At long wavelengths the plateau of
the uncoupled instability is still apparent (provided
$d\Sigma/d\gdot<0$) and unperturbed; for $d\Sigma/d\gdot>0$ this
plateau  disappears to leave only the diffusive,
concentration-coupled bump. The dispersion relation for a system
closer to type II ($D$ reduced by a factor 100, at fixed $G'/\zeta$)
is shown in Fig.~\ref{fig:dispersion_phi0.11:f}: the enhancement at
long length scales is much more pronounced, corresponding to the
greater spinodal shift (triangles of
Fig.~\ref{fig:spinodals_with_phi}).  However the mechanical plateau
(present when $d\Sigma/d\gdot<0$) is still unperturbed at long length
scales (though indiscernible on the scale of
Fig.~\ref{fig:dispersion_phi0.11:f}).

The overall dispersion shape (the same in type I and II systems) is
captured by analysing the simplified stability
matrix~\ref{eqn:fluct_in_y_matrix_conc}. We consider two separate
cases:

\begin{description}

\item[(a) ] 
  
  Shear rates {\em above} the lower spinodal of the {\em coupled}
  model but which are stable in the uncoupled limit
  ($d\Sigma/d\gdot>0$;
  Fig.~\ref{fig:dispersion_coupled_phi0.11_gdot2.0_and_4.0_2D}, left).
  Here, we find the following regimes:

\begin{description}

\item[(i)]
\underline{Diffusive regime $k\to 0$} in which
\be
\label{eqn:dispersion_coupled_regime1}
\omega=-\left[\tilde{D}+\frac{1}{\tilde{\zeta}}\frac{\df}{\djs}\right]k^2.
\ee
This is marked as a dashed line in Fig.
~\ref{fig:dispersion_coupled_phi0.11_gdot2.0_and_4.0_2D}(left), and
underestimates the exact result because it neglects feedback via the
elastic free energy $F^{\rm e}$~\cite{neglect_Y}. The growth rate in
this regime is limited by the rate at which matter diffuses a distance
$O(1/k)$: momentum diffusion and micellar strain response are
adiabatic in comparison.  Note that for larger shear rates for which
$d\Sigma/d\gdot<0$ (discussed in {\bf (b)}, below)
Eqn.~\ref{eqn:dispersion_coupled_regime1} is negative, so this branch
is absent from the instability (compare
Figs.~\ref{fig:dispersion_coupled_phi0.11_gdot2.0_and_4.0_2D} left and
right).

\item[(ii)]
  
  \underline{Non-conserved ``plateau'' regime.} For larger $k$, the
  rate at which the non-conserved micellar strain can respond (even
  within concentration enhanced dynamics) is the limiting factor;
  concentration diffusion becomes adiabatic in comparison. If the
  eventual interfacial cutoff in the dispersion relation once
  $l^2k^2=O(1)$ or $\xi^2 k^2=O(1)$ were absent we would then see a
  non-conserved $k-$independent plateau regime in which 
\be
\label{eqn:dispersion_coupled_regime2}
\omega_{\rm pl}=\frac{\tdjs +\frac{\D\tdf}{\D
    \tilde{D}\tilde{\zeta}}}{1+\bar{Z}-\frac{\D
    G'b \bar{W}_{xy}^2}{\D\tilde{D}\tilde{\zeta}}}, \ee
with
\be
\tdi=\di \frac{\eta_s\tau_d}{k^2},\quad i\in {F,M}.
\ee
However, for the systems of interest to us the low-$k$ crossover to
this regime is not well separated from the interfacial cutoff and the
plateau is replaced by a rounded maximum at $(k^*, \omega^*)$ (thus
defined) where $\omega^*\lae \omega_{\rm pl}$.  This maximum selects
a length scale $k^{*-1}$.

\item[(iii)]
  
  \underline{High $k$ interfacial cutoff.} The dispersion relation is
  cut off by interfaces once $k^2l^2=O(1)$ or $k^2\xi^2=O(1)$. $l$ and
  $\xi$ are of similar order for the systems of interest to us.

\end{description}

An estimate for the selected wavevector $k^*$ can be be obtained by
expanding about $\omega\approx \omega_{\rm pl}$ to
find
\be
\label{eqn:dispersion_peak}
k^{*4}\approx\frac{\omega_{\rm pl} }{\tilde{D}\xi^2  -\frac{\D\tilde{D}(1+\bar{Z})l^2}{\D\hdjs} -\frac{\D G'b\bar{W}_{xy}^2l^2}{\D\tilde{\zeta}\hdjs}},
\ee
where
\be
\hdjs=\tdjs-\omega_{\rm pl}(1+\bar{Z})<0.
\ee
This $k^*$ is marked by a dashed arrow in
Fig.~\ref{fig:dispersion_coupled_phi0.11_gdot2.0_and_4.0_2D}, and
agrees reasonably with the numerics. As in the conventional CH
instability, $k^*\to 0$ at the spinodal boundaries (where $\omega^*\to
0$).  This is not visible in figures~\ref{fig:dispersion_phi0.11}
and~\ref{fig:dispersion_phi0.02}, because $k^*$ only starts to
diminish appreciably for indiscernibly small $\omega^*$ on our scale.
Note that Eqn.~\ref{eqn:dispersion_peak} doesn't reproduce the
selected wavevector of standard CH theory in zero shear, since phase
separation is still affected by coupling of composition to
viscoelastic effects~\cite{OnuTan97} even in this limit.

\item[(b)]
  
  For higher shear rates that would have been unstable even in the
  uncoupled limit, the dispersion relation develops a shoulder at
  small $k$: see
  Fig.~\ref{fig:dispersion_coupled_phi0.11_gdot2.0_and_4.0_2D}(right).
  As noted above, this is just the large length scale part of the pure
  mechanical dispersion branch (Sec.~\ref{sec:uncoupled_fc}),
  comprising a Reynolds regime and a mechanical non-conserved regime.
  (See regimes $\bf (i)$ and $\bf (ii)$ in
  Fig.~\ref{fig:dispersion_coupled_phi0.11_gdot2.0_and_4.0_2D}(right).)
  The growth rate here is much faster than diffusion so concentration
  is absent from the eigenvector.  At shorter length scales,
  concentration {\em can} keep pace and {\em is} included. For shear
  rates that are not too deep inside the unstable region, the
  dispersion relation then rises to the rounded plateau estimated by
  Eqn.~\ref{eqn:dispersion_coupled_regime2} (regime $\bf (iii)$ of
  Fig.~\ref{fig:dispersion_coupled_phi0.11_gdot2.0_and_4.0_2D}(right))
  before finally being cut off by interfaces (regime $\bf (iv)$). The
  maximum at $k^*$ is again estimated by
  Eqn.~\ref{eqn:dispersion_peak} (marked by the dashed arrow in
  Fig.~\ref{fig:dispersion_coupled_phi0.11_gdot2.0_and_4.0_2D}(right)).
 
\end{description}

The preceding analysis captures the qualitative features of the
dispersion relations in many regimes.
However, some more exotic effects are apparent in
Figs~\ref{fig:dispersion_phi0.11:c} and~\ref{fig:dispersion_phi0.11:e}
for shear rates well above the lower spinodal. For $20\lae \gdot\lae
80$, concentration coupling gives {\em negative} feedback at short
length scales. The origin of this (not included in our above
analytical treatment) is that the velocity advecting the micellar
backbone strain is not the centre of mass velocity $\vect{v}$ (as the
above analytical work assumed) but the micellar velocity
$\vm=\vect{v}+(1-\phi)\vrel$. A fluctuation $\delta W_{yy}$ in general
causes a fluctuation in $\phi$, and therefore in $\vrel$. When
included in the advective term, this feeds back negatively on
$W_{yy}$.  At still higher shear rates $\gdot>80$ in
Fig.~\ref{fig:dispersion_phi0.11:e}, the dispersion relation has a
pronounced ridge corresponding to the high shear rate lobe discussed
above and schematised by the right hand dashed line of
Fig.~\ref{fig:cartoon}b.

\subsubsection{Fluctuations in the vorticity direction}
\label{sec:vorticity}

In the uncoupled limit $\zeta\to\infty$, the mechanical subspace is
stable with respect to vorticity fluctuations at all shear rates,
while concentration has the usual CH demixing instability for $D<0$.
Can coupling influence this instability?  In some works
\cite{milner93,schmitt95} spinodal shifts have indeed been observed.
In our model this does not occur, for the following reason.  By
analogy with the feedback mechanism studied above for
$\vect{k}=k\hat{\vect{y}}$, the term in Eqn.~\ref{eqn:concentration}
that could participate in positive feedback is
$\bar{W}_{zz}G'(\phi)k^2\delta \phi$.  In our model
(unlike~\cite{milner93,schmitt95}) $\bar{W}_{zz}=0$
(Eqn.~\ref{eqn:intrinsic_flow_curve}) so the stability of vorticity
fluctuations is unaffected by shear.  Accordingly, hereafter we
consider only $\vect{k}=k\hat{\vect{y}}$.

\section{Shear startup experiment}
\label{sec:startup}

\subsection{Time-dependence and linear analysis}

The stability analysis of startup flow is more involved, because here
fluctuations emerge against a background state that itself evolves,
deterministically, in time.  We first outline these deterministic
kinetics (for an idealised noiseless system) before analysing
fluctuations.

\subsubsection{Deterministic ``background'' kinetics}

\begin{figure}[h]
\centerline{\psfig{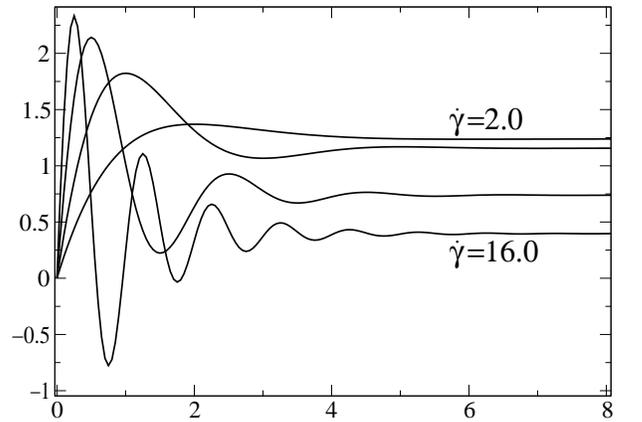}}
\caption{Homogeneous background micellar strain $\bar{W}_{xy}$ {\it vs.} $t$ for $\gdot=2.0,\,4.0,\,8.0,\,16.0$ (top to bottom at right of plot).
\label{fig:background_stress} } 
\end{figure}

At time $t=0$, the rheometer plate at $y=L$ is set in motion with
velocity $\bar{\gdot}L\hat{\vect{x}}$, giving an instantaneous shear
rate profile $\gdot(y,0)=\bar{\gdot}\delta(y-L)$. Without noise, the
ultimate steady state would be homogeneous. Firstly, on the Reynolds
time scale $\tau_{\rm R}=\rho L^2/\eta$ the shear rate rapidly
homogenizes across the cell such that $\gdot(y)=\bar{\gdot}$.
Secondly, on the Maxwell time scale $\tau\gg\tau_{\rm R}$ the micellar
strain starts to evolve homogeneously, according to
Eqn.~\ref{eqn:JSd}, as
\bea
\label{eqn:background_time_ev1}
W_{xy}(t)&=&\frac{\bgdot}{1+b\bgdot^2}\left\{1-e^{-t}\left[\cos(\sqrt{b}\bgdot t)-\sqrt{b}\bgdot \sin(\sqrt{b}\bgdot t)\right]\right\},\nonumber\\
W_{yy}(t)&=&-\frac{1}{1+a}\frac{b\bgdot^2}{1+b\bgdot^2}\nonumber\\
         & &\times\left\{1-e^{-t}\left[\cos(\sqrt{b}\bgdot t)+\frac{1}{\sqrt{b}\bgdot}\sin(\sqrt{b}\bgdot t)\right]\right\},\nonumber\\
W_{xx}(t)&=&\frac{1+a}{a-1}W_{yy}(t),\nonumber\\
W_{zz}(t)&=&W_{xz}(t)=W_{yz}(t)=0
\eea 
(see Fig.~\ref{fig:background_stress}). Although these expressions
recover Eqn.~\ref{eqn:intrinsic_flow_curve} as $t\to\infty$ (so that
the total shear stress would then be on the intrinsic constitutive
curve), we show below that in general the flow goes unstable before
this limit is reached.

\subsubsection{Inhomogeneous fluctuations}

In a real system, these homogeneous transients represent only a
background state
$\vect{\bar{u}}(t)=[\bar{\gdot},\bar{\tens{W}}(t),\bar{\phi}]$,
which is subject to fluctuations:
\be 
\label{eqn:linearize_startup}
\left( \begin{array}{c} \gdot(\vect{r},t) \\[5truept] \tens{W}(\vect{r},t)
    \\[5truept] \phi(\vect{r},t) \end{array} \right) 
   = \left( \begin{array}{c}  \bar{\gdot} \\[5truept] \bar{\tens{W}}(t) \\[5truept] \bar{\phi} \end{array} \right)
   + \sum_{\vect{k}} \left( \begin{array}{c} \delta \gdot(t) \\[5truept] \delta \tens{W}(t) \\[5truept] \delta \phi(t) \end{array} \right)_{\vect{k}}e^{i\vect{k}.\vect{r}}. \ee
   To investigate the stability of these fluctuations, we linearize
   the model's dynamical equations
   (\ref{eqn:navier},~\ref{eqn:incomp},~\ref{eqn:concentration},~(\ref{eqn:JSd})
   to get
\be
\label{eqn:langevin_startup}
\partial_t\delta \vect{u}_{\vect{k}}(t)=\tens{M}_{\vect{k}}(t).\delta\vect{u}_{\vect{k}}(t).
\ee
This is the counterpart in startup of Eqn.~\ref{eqn:langevin}, with
the important new feature that $\tens{M}_{\vect{k}}(t)$ is time
dependent, via its dependence on the homogeneous background state
$\bar{u}(t)=[\bar{\gdot},\bar{\tens{W}}(t),\bar{\phi}]$ and hence on the
evolution of the micellar strain $\bar{\tens{W}}$ {\em towards}
the intrinsic constitutive curve.
The eigenmodes are therefore now time-dependent:
\be
\label{eqn:time_dep_vectors}
\omega_{\vect{k},\alpha}(t)\eigenvec(t)=\tens{M}_{\vect{k}}(t)\eigenvec(t).
\ee

In any startup experiment, then, the micellar strain evolves over a
time $\tau_{\rm ss}=O(\tau)$ (thus defined) towards the intrinsic constitutive
curve, as described above. The dispersion relation
$\omega_{\vect{k},\alpha}(t)$ correspondingly evolves towards the one
given by Eqn.~\ref{eqn:find_modes} for an initial condition on that
flow curve. So for a shear rate in the unstable region, at least one
dispersion branch must go positive at some time $t_0\le\tau_{\rm ss}$
so that the homogeneous transient
$[\bar{\gdot},\bar{\tens{W}}(t),\bar{\phi}]$ goes unstable.  In most
regimes we find only one positive branch~\cite{multi_branch}  and drop the ``mode'' subscript $\alpha$, with the
understanding that we mean  the largest branch.  At wavevector
$\vect{k}$, the amplitude of the growing fluctuations at a time $t>t_0$ is
approximately set by
\be
A(\vect{k},t)\sim \exp\left[\int_{t_0}^{t} dt' \omega_{\vect{k}}(t')\right].
\ee
We choose a rough criterion for detectability by scattering to be $\log
A=O(10)$. This defines a wavevector-dependent time scale
$\tau_{\rm inst}(\vect{k})$, via
\be
\label{eqn:condition_for_inst}
\int_{t_0}^{\tau_{\rm inst}(\vect{k})} dt' \omega_{\vect{k}}(t')=O(10).
\ee
In most regimes, there is a selected wavevector $\vect{k}^*$ at which
fluctuations emerge fastest, as the result of a peak in the dispersion
relation $\omega_{\vect{k}}(t)$ {\it vs.}  $\vect{k}$. In practice,
the peak shifts along the $k$ axis in time, but it is still usually
possible to obtain a reasonable estimate of the overall $k^*$; we
justify this claim below. We therefore define the overall time scale
of instability to be
\be
\label{eqn:condition_for_inst_overall}
\tau_{\rm inst}=\tau_{\rm inst}(\vect{k}^*).
\ee

\begin{figure}[h]
\centerline{\psfig{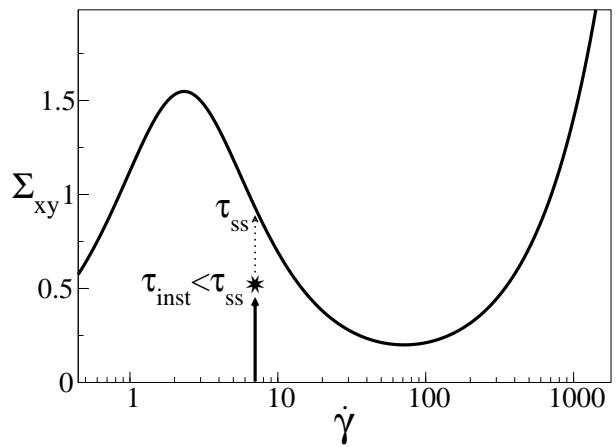}}
\caption{Cartoon:  homogeneous startup flow going unstable at time $\tau_{\rm inst}$ before it can reach   the intrinsic unstable constitutive curve at time $\tau_{\rm ss}$.
\label{fig:schematic_stress_evolution} } 
\end{figure}

By the time $\tau_{\rm inst}$, then, the system is measurably
inhomogeneous, and our linear calculation terminates.  In general,
this occurs well before the intrinsic constitutive curve would have been
attained, \ie\, $\tau_{\rm inst}<\tau_{\rm ss}$
(Fig.~\ref{fig:schematic_stress_evolution}), so that the instability is
determined not by the time-independent dispersion relations of
Sec.~\ref{sec:onflowcurve} above, but by their time-dependent
counterparts (given below).

Is the unstable intrinsic constitutive curve ever attained before the
instability occurs, such that $\tau_{\rm ss}\ll\tau_{\rm inst}$? A
{\em necessary} condition is that the growth rate $\omega_{\rm
  fc}=\omega_{\vect{k}^*}(t=\tau_{\rm ss})$ that would occur once the
flow curve were reached (given by the dispersion relations of
section~\ref{sec:onflowcurve}) obeys
\be
\label{eqn:simple}
\omega_{\rm fc}\tau_{\rm ss}\ll 1.
\ee
This is {\em not} usually satisfied (recall
Figs.~\ref{fig:dispersion_phi0.11} and~\ref{fig:dispersion_phi0.02})
since $\omega_{\rm fc}$ is itself set by the Maxwell time $\tau$ (with
a prefactor set by the slope of the flow curve and by concentration
coupling). Nonetheless, condition~\ref{eqn:simple} is satisfied just
inside the spinodal, since $\omega_{\rm fc}\to 0$ smoothly at the
spinodal.  However, this condition is not always {\em sufficient}.  In
particular, for shear rates just inside the {\em upper} spinodal, the
homogeneous micellar strain oscillates strongly in startup.
Correspondingly, the growth rate significantly overshoots $\omega_{\rm
  fc}$ (Figs.~\ref{fig:startup_raw} to~\ref{fig:startup_D-9} below)
and fluctuations still emerge before the intrinsic constitutive curve would be
attained. In fact, these oscillations mean that fluctuations can
become (temporarily) unstable in startup, even for shear rates above
the upper spinodal (as defined via slow shear rate sweeps).  This upper
spinodal is therefore not particularly relevant to
startup flows.

For shear rates just inside the {\em lower} spinodal,
condition~\ref{eqn:simple} is necessary {\em and} sufficient, and the
intrinsic constitutive curve {\em is} then attained before the instability
develops appreciably.  Here we can assume that the stability matrix
changes discontinuously at $t=0$ from the stable
$\tens{M}_{\vect{k}}(t=0)$ (with $\bar{\gdot}=\bar{\tens{W}}=0$), to
the unstable matrix $\tens{M}_{\vect{k}}(t=\infty)$ for a state on the
intrinsic constitutive curve. The instability is then, even in startup,
determined by the time-independent dispersion relations of
Sec.~\ref{sec:onflowcurve}.

We pause to compare our analysis with that of Cahn and Hilliard for a
two component system temperature-quenched at time $t=0$ into the
unstable region, $\partial \mu(\phi,T)/\partial \phi <0$. A good
approximation, invariably made, is that $\mu(\phi)$ changes
discontinuously at $t=0$ from its initial stable state to the final
one of negative slope, \ie\ that the heat diffuses out instantaneously
with respect to the time scale at which fluctuations grow.  We have
just seen that the corresponding assumption for our purposes (the
background state $\bar{u}(t)$ instantaneously reaching the intrinsic
constitutive curve) is not in general valid.

In the next section we present results for the time-dependent unstable
dispersion branch over the time interval $t_0\to \tau_{\rm inst}$ for
several startup quenches, indicating in each case the selected
wavevector $k^*$. We also give results for the time dependent
eigenvector (at $k^*$) noting whether separation occurs predominantly
in the mechanical variables or in concentration.

\subsection{Results: uncoupled model}

\begin{figure*}[htbp]
\centering
\subfigure[$\gdot=7.0$, dispersion relation.]{
\includegraphics[scale=0.34]{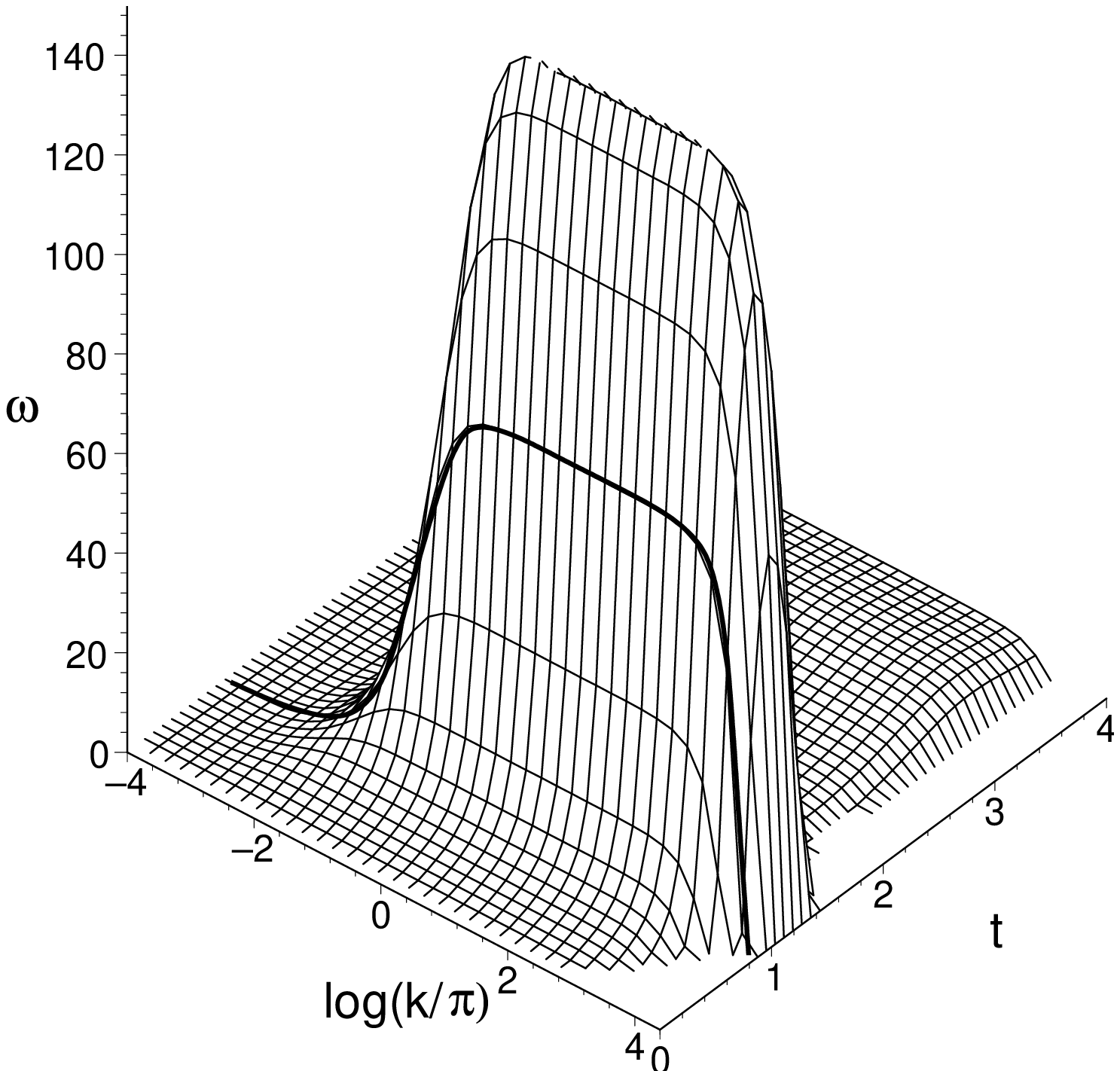}
\label{fig:startup_raw:a}
}
\subfigure[$\gdot=40.0$, dispersion relation.]{
\includegraphics[scale=0.34]{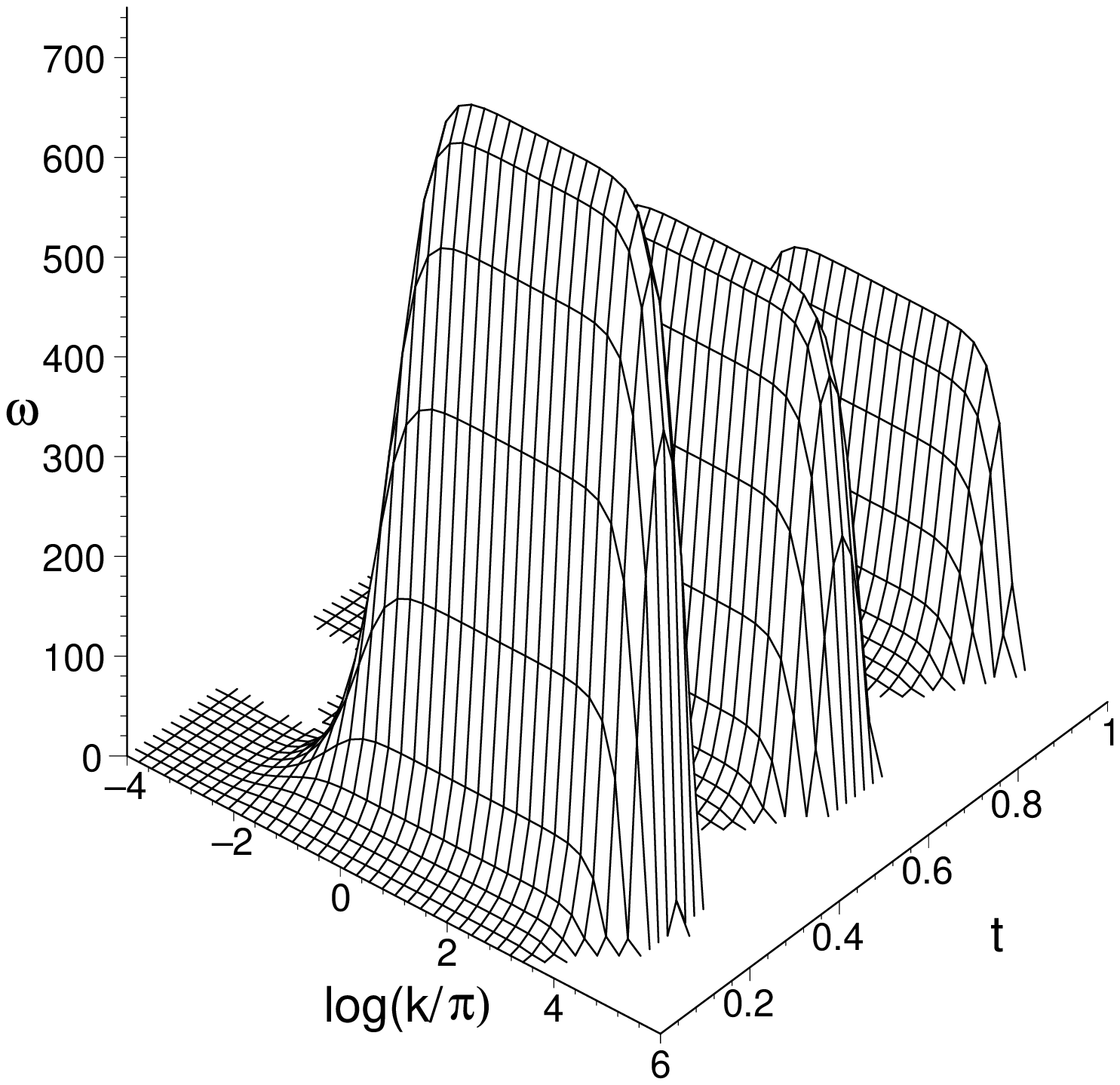}
\label{fig:startup_raw:c}
}
\subfigure[$\gdot=70.0$, dispersion relation]{
\includegraphics[scale=0.34]{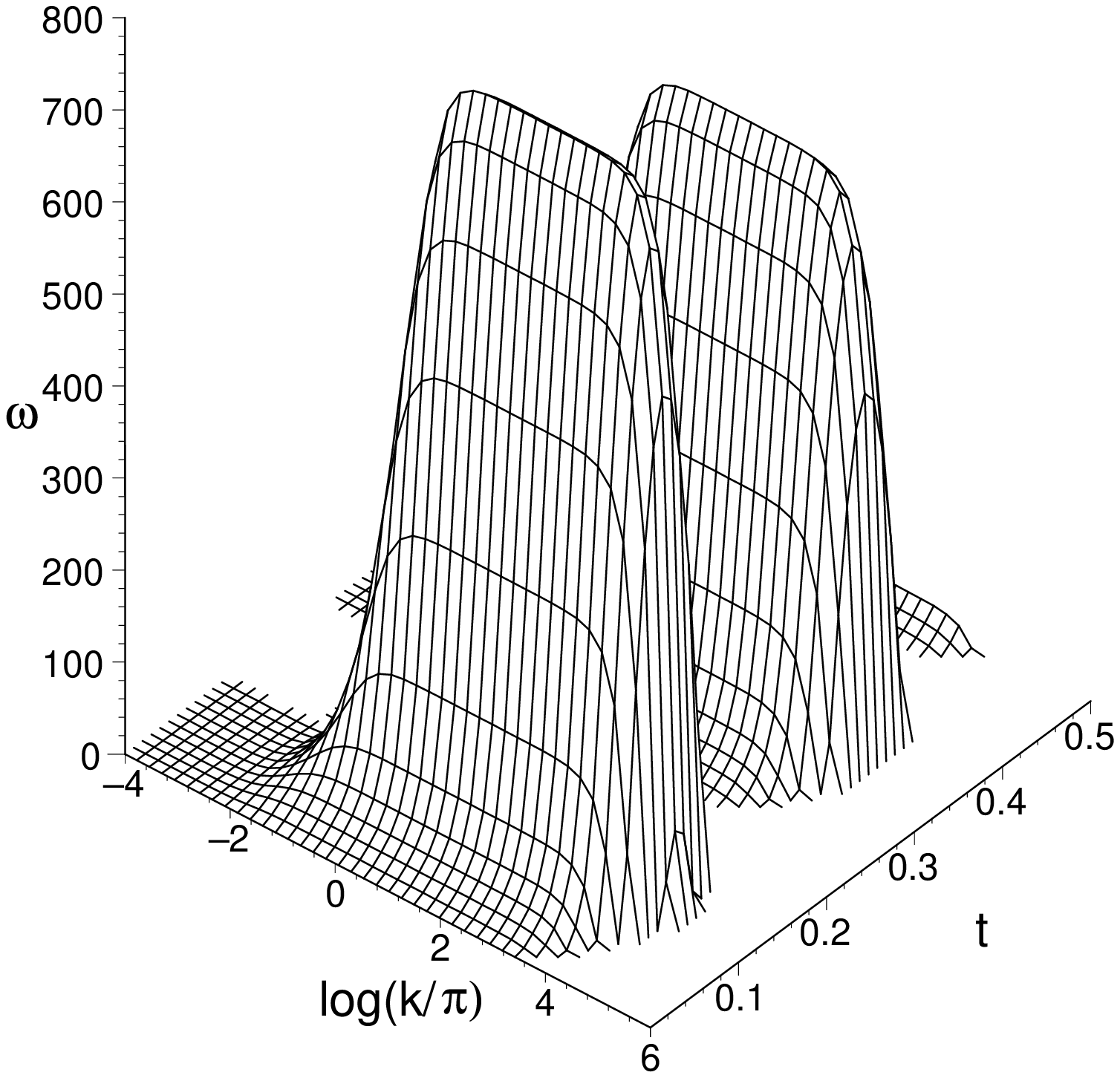}
\label{fig:startup_raw:e}
}
\subfigure[$\gdot=7.0$, eigenvector at $\log(k^*/\pi)=0.0$.]{
  \includegraphics[scale=0.20]{vecmax_t_raw_phi0.11_gdot7.0.eps}
\label{fig:startup_raw:b}
}
\subfigure[$\gdot=40.0$, eigenvector at $\log(k^*/\pi)=0.0$.]{
  \includegraphics[scale=0.20]{vecmax_t_raw_phi0.11_gdot40.0.eps}
\label{fig:startup_raw:d}
}
\subfigure[$\gdot=70.0$, eigenvector at $\log(k^*/\pi)=0.0$.]{
  \includegraphics[scale=0.20]{vecmax_t_raw_phi0.11_gdot70.0.eps}
\label{fig:startup_raw:f}
}
\caption{Type A instabilities in a type I system: time dependent dispersion relation (top) and eigenvector (bottom) in the uncoupled limit $\zeta\to\infty$ for $\phi=0.11$. The rheological model parameters all assume the experimental values of table~\ref{table:parameters}. The thick line in Fig.~\ref{fig:startup_raw:a} and the arrows in Figs. d,e,f denote the time at which the instability becomes measurable.  The discontinuities in the first derivative of the eigenvector components result from a crossing of eigenvalues, discussed in the text.}
\label{fig:startup_raw}. 
\end{figure*}

Fig.~\ref{fig:startup_raw} (top) shows the numerically calculated
startup dispersion relation $\omega_{\vect{k}}(t)$ in this uncoupled
limit $\zeta\to \infty$. The oscillations arise from the oscillations in
$\bar{\tens{W}}(t)$ towards the intrinsic constitutive curve
(Fig.~\ref{fig:background_stress}).  Despite the time dependence, the
main features of the time-independent dispersion relation for
fluctuations about the intrinsic constitutive curve
(Fig.~\ref{fig:dispersion_phi0.11:a}) are still apparent: there is a
Reynolds regime as $k\to 0$, a non-conserved plateau regime at
intermediate $k$, and interfacial cutoff at large $k$.  As before,
then, in this uncoupled limit there is no  selected
wavevector $k^*$.

For each startup, we estimated the time $\tau_{\rm inst}$ at which the
instability would become measurable, as governed by
criterion~\ref{eqn:condition_for_inst} applied to wavevectors in the
plateau regime.  It is marked by the thick line in
Fig.~\ref{fig:startup_raw:a} and an arrow in
Figs.~\ref{fig:startup_raw:b} to~\ref{fig:startup_raw:f}. For each
value of $\gdot$ in Fig.~\ref{fig:startup_raw}, we find $\tau_{\rm
  inst}\ll \tau_{\rm ss}$: instability occurs long before the
underlying flow curve would have been attained. 

Fig.~\ref{fig:startup_raw} (bottom), shows the time-dependent
eigenvector at wavevector $k^*=\pi$ (chosen arbitrarily since the
eigenvector is independent of $k$ in the plateau regime). This is
dominated by $\delta \gdot$, since $\delta W_{xy}+\eta\delta\gdot =0$
in this zero-Reynolds regime. Note also that the normal stress,
encoded in $\delta Z$, dominates the shear contribution $\delta
W_{xy}$: consistently with the remarks of
section~\ref{sec:uncoupled_spinodal}, the normal stress plays an
important role in this mechanical instability.

The discontinuity in the first derivative of eigenvector is due to a
crossing of two positive eigenvalues: in contrast to the
time-independent dispersion relations for fluctuations about the
intrinsic constitutive curve, in startup there is sometimes more than
one positive dispersion branch, the smaller one of which can be in the
subspace $[ik\delta v_z,W_{xz},W_{yz}]$.  {\em However}, this second
unstable mode only occurs at high shear rates $\gdot\gae 10$, and even
then only crosses the first for times well after $\tau_{\rm inst}$:
consistent with the claim made above, we never observe mode crossing
in the relevant time regime $t\le\tau_{\rm inst}$. This also applies
to the coupled model, to which we now turn.

\begin{figure*}[htbp]
\subfigure[$\gdot=2.0$, dispersion relation.]{
    \includegraphics[scale=0.34]{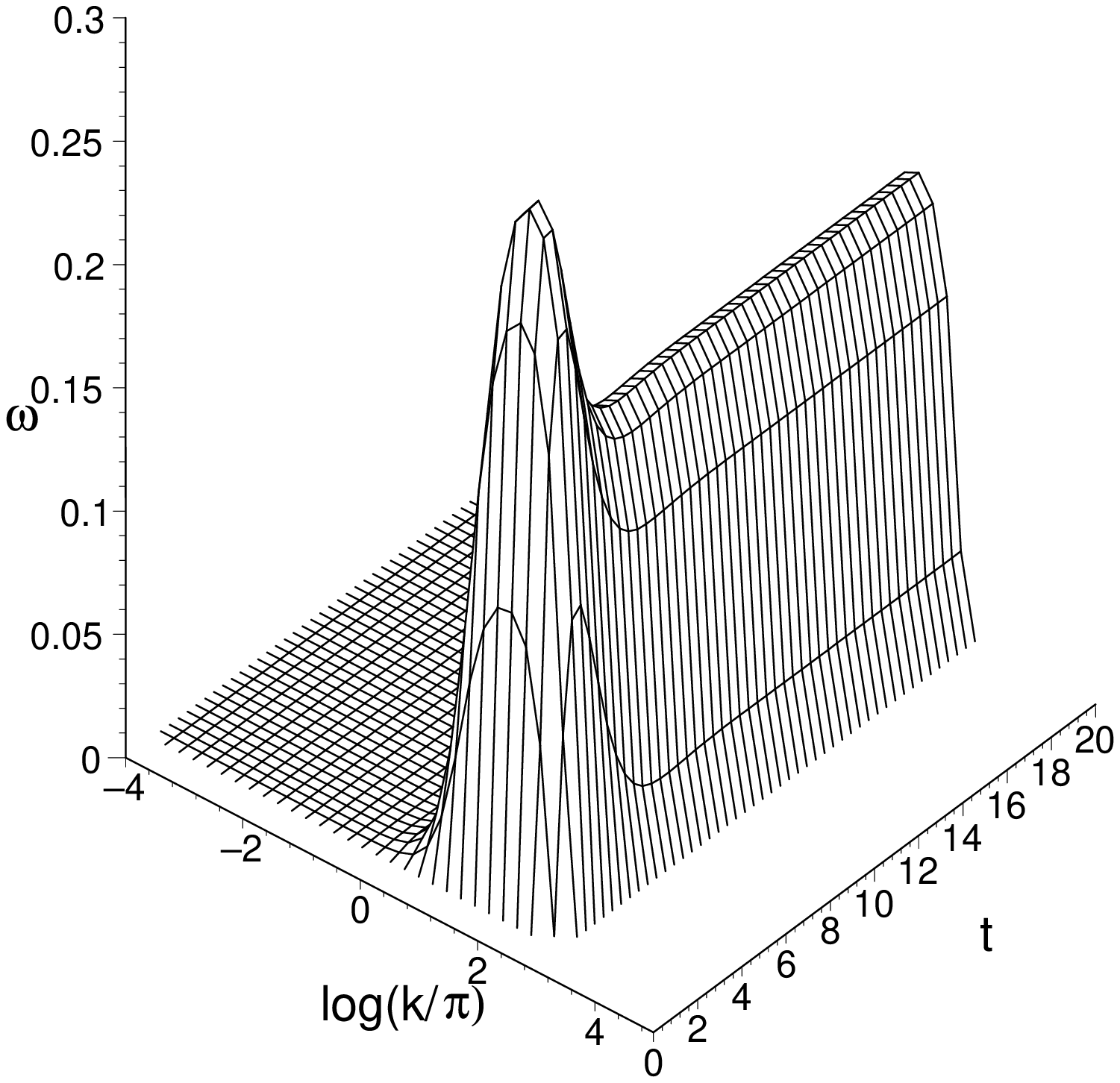}
\label{fig:startup_D-4:a}
}
\subfigure[$\gdot=4.0$, dispersion relation.]{
\includegraphics[scale=0.34]{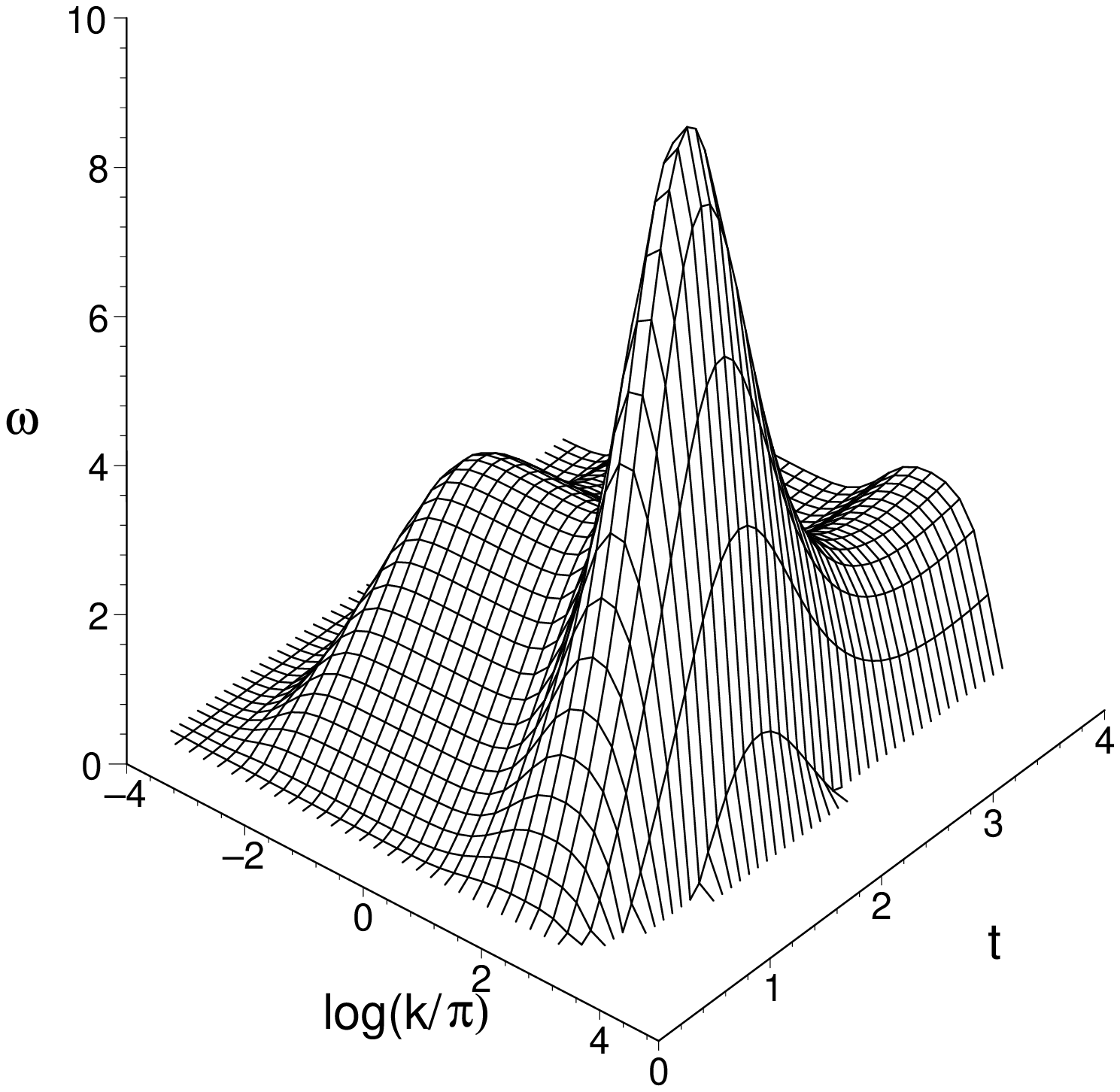}
\label{fig:startup_D-4:c}
}
\subfigure[$\gdot=35.0$, dispersion relation.]{
\includegraphics[scale=0.34]{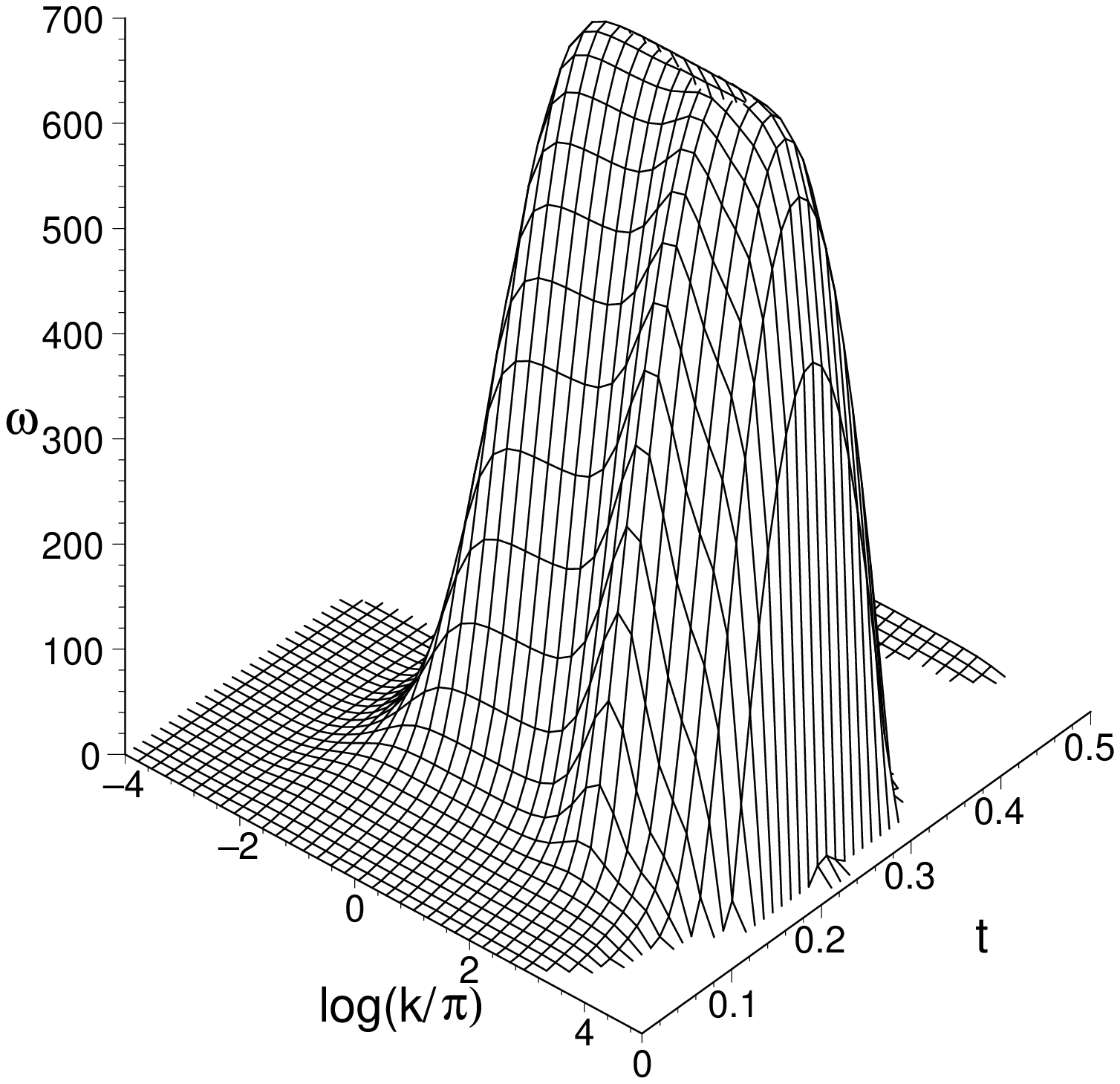}
\label{fig:startup_D-4:e}
}
\subfigure[$\gdot=2.0$, eigenvector at $\log(k^*/\pi)=2.0$.]{
  \includegraphics[scale=0.20]{vecmax_t_D2.6e-4_phi0.11_gdot2.0.eps}
\label{fig:startup_D-4:b}
}
\subfigure[$\gdot=4.0$, eigenvector at $\log(k^*/\pi)=2.0$.]{
  \includegraphics[scale=0.20]{vecmax_t_D2.6e-4_phi0.11_gdot4.0.eps}
\label{fig:startup_D-4:d}
}
\subfigure[$\gdot=35.0$, eigenvector at $\log(k^*/\pi)=2.4$.]{
  \includegraphics[scale=0.20]{vecmax_t_D2.6e-4_phi0.11_gdot35.0.eps}
\label{fig:startup_D-4:f}
}
\caption{Type A instabilities in a type I system: time dependent dispersion relation (top) and eigenvector (bottom) for a coupled model in which all parameters assume the experimental values of table~\ref{table:parameters}. The concentration $\phi=0.11$. The arrows in Figs. e,f show the time at which the instability first becomes measureable (the instability occurs beyond the time window of Fig. d). The discontinuities in the first derivative of the eigenvector components  result from a crossing of eigenvalues, discussed in the text. The instability time $\tau_{\rm inst}$ occurs beyond the displayed time window for $\gdot=2.0$.}
\label{fig:startup_D-4}. 
\end{figure*}

\begin{figure*}[htbp]
  \centering \subfigure[$\gdot=0.01$, dispersion relation.]{
    \includegraphics[scale=0.34]{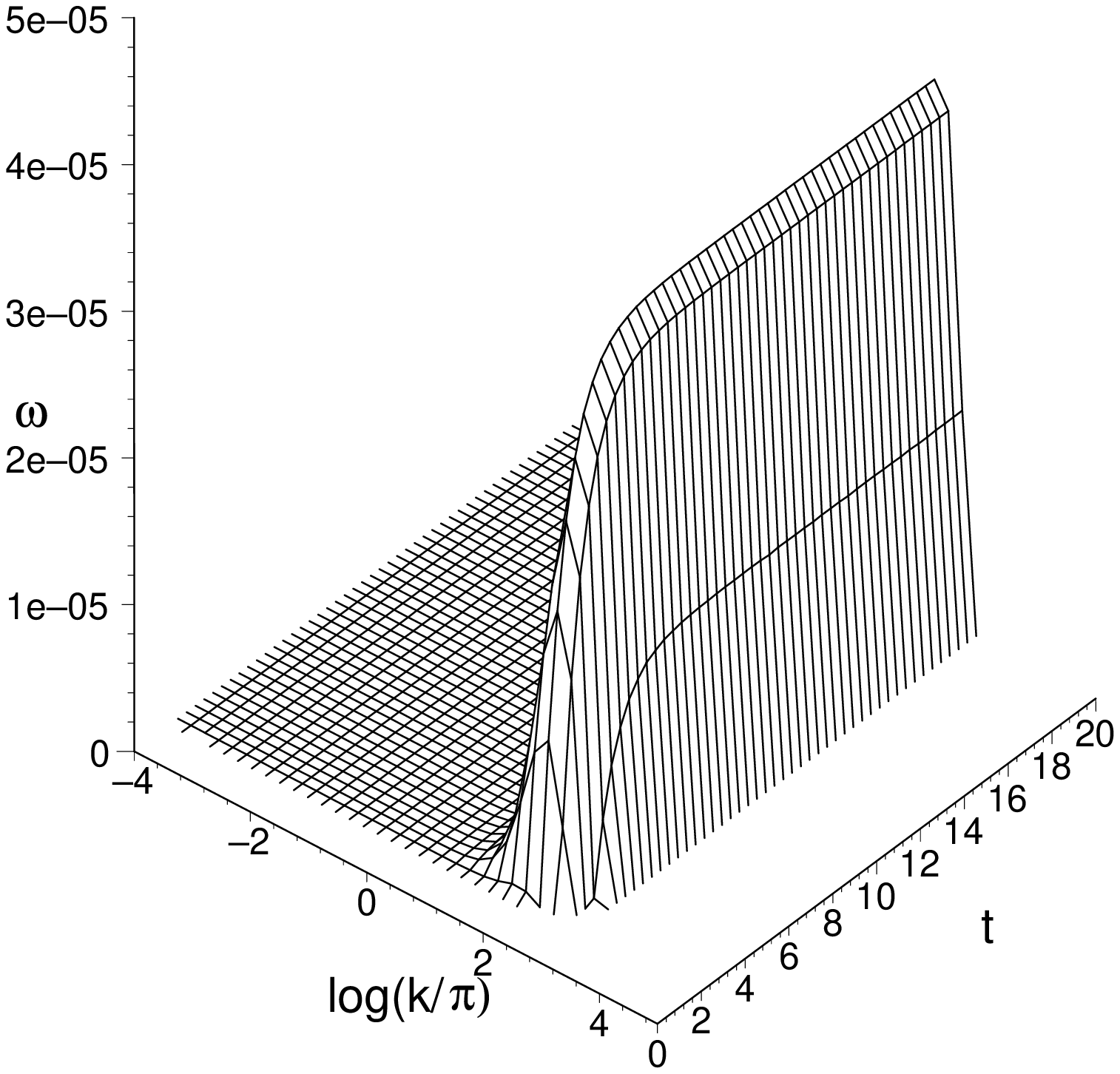}
\label{fig:startup_D-9:a}
}
\subfigure[$\gdot=1.0$, dispersion relation.]{
\includegraphics[scale=0.34]{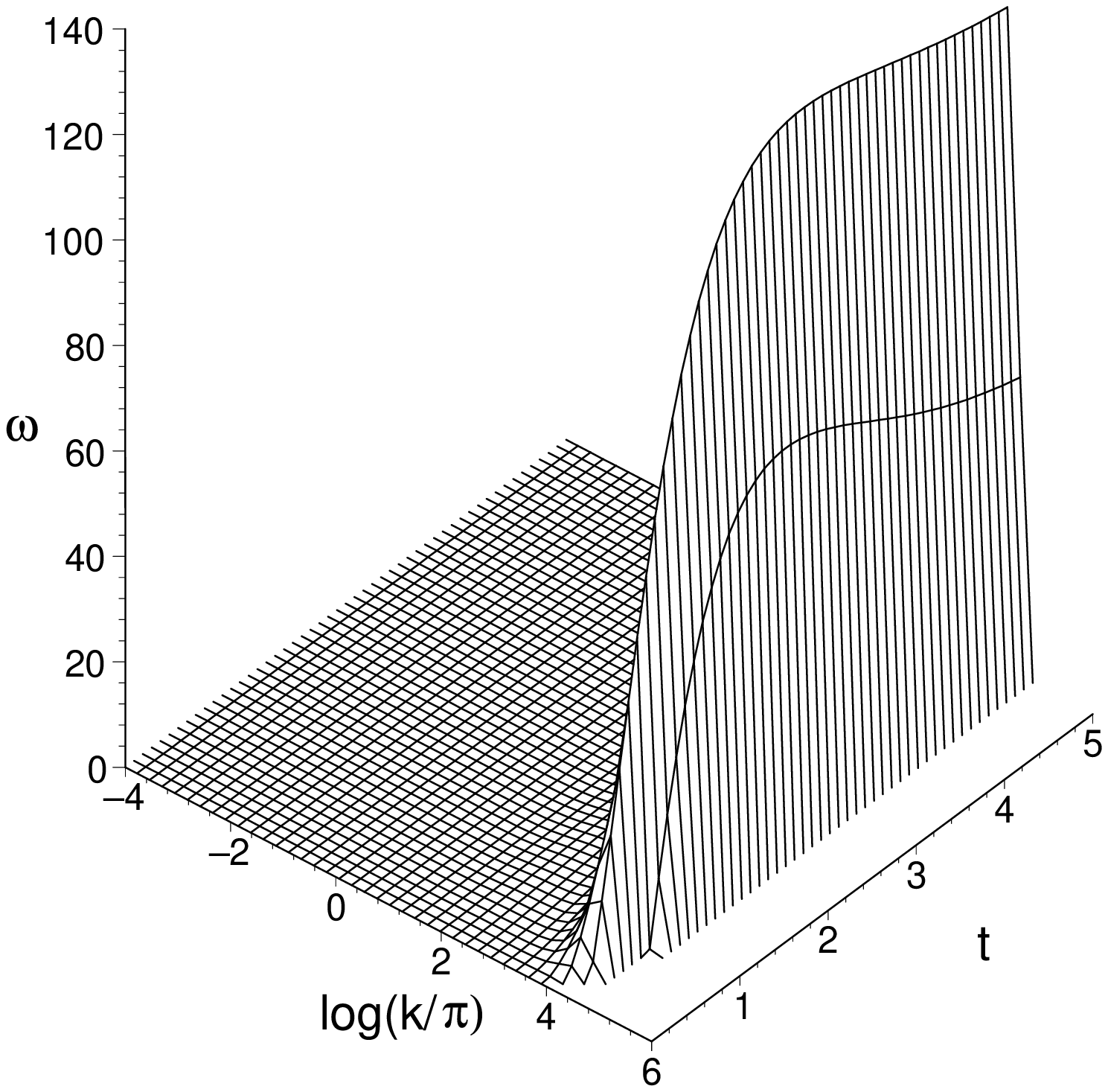}
\label{fig:startup_D-9:c}
}
\subfigure[$\gdot=10.0$, dispersion relation.]{
\includegraphics[scale=0.34]{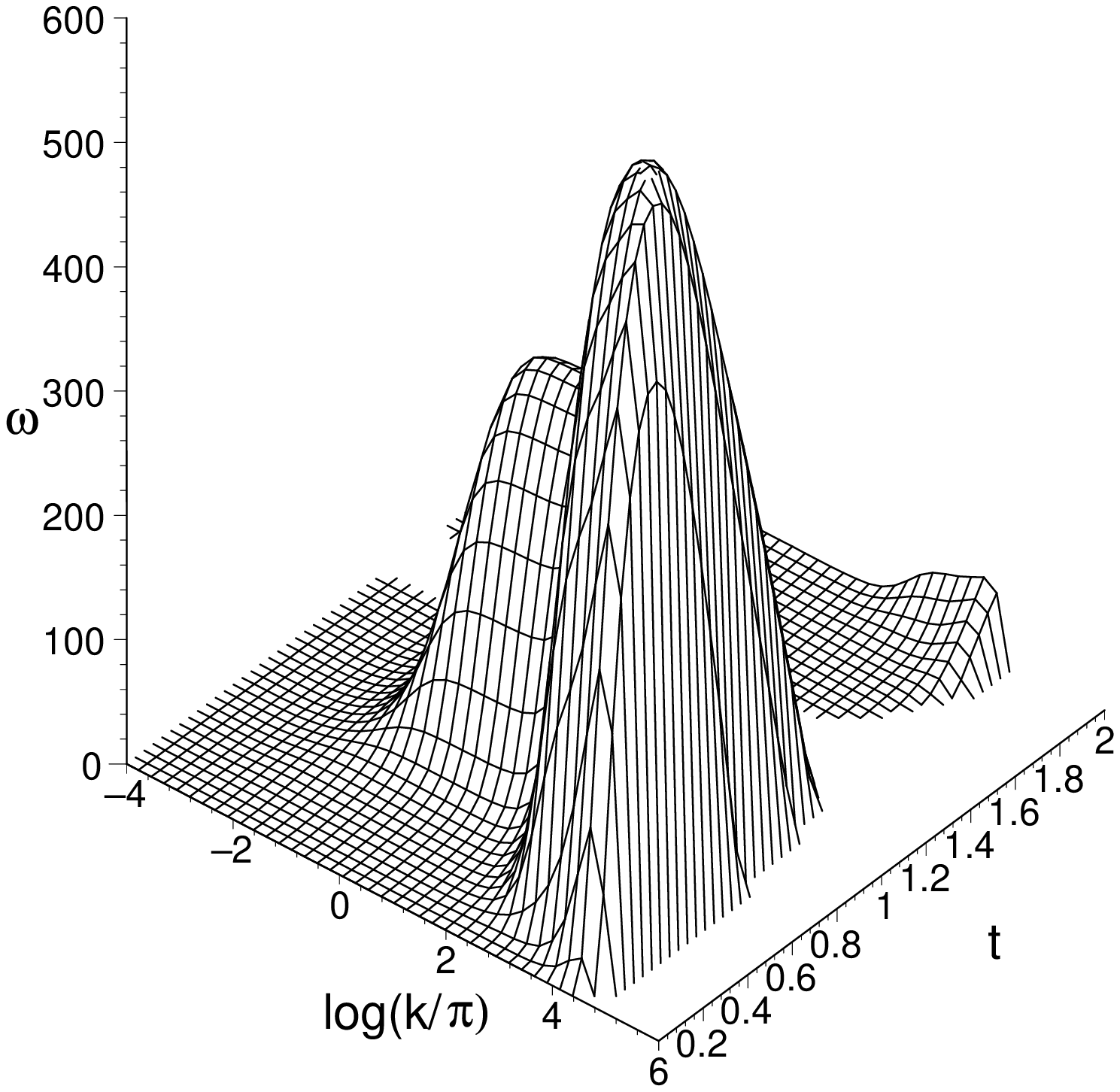}
\label{fig:startup_D-9:e}
}
\subfigure[$\gdot=0.01$, eigenvector at $\log(k^*/\pi)=2.4$.]{
  \includegraphics[scale=0.20]{vecmax_t_D2.6e-9_phi0.11_gdot0.01.eps}
\label{fig:startup_D-9:b}
}
\subfigure[$\gdot=1.0$, eigenvector at $\log(k^*/\pi)=4.4$.]{
  \includegraphics[scale=0.20]{vecmax_t_D2.6e-9_phi0.11_gdot1.0.eps}
\label{fig:startup_D-9:d}
}
\subfigure[$\gdot=10.0$, eigenvector at $\log(k^*/\pi)=4.4$.]{
  \includegraphics[scale=0.20]{vecmax_t_D2.6e-9_phi0.11_gdot10.0.eps}
\label{fig:startup_D-9:f}
}
\caption{Type A instabilities (a,b,d,e) and type B instability (c,f) in a type II system: time dependent dispersion relation (top) and eigenvector (bottom) for a coupled model in which all parameters assume the experimental values of table~\ref{table:parameters}, {\em except} $D$, which is reduced by a factor $10^5$. The concentration $\phi=0.11$. The arrows in Figs. e,f show the time at which the instability first becomes measureable (the instability occurs beyond the time window of Fig. d).}
\label{fig:startup_D-9}. 
\end{figure*}

\subsection{Results: coupled model}

We now give startup results for the coupled model. Denoting the
experimental DLS value of the diffusion coefficient $D$
(table~\ref{table:parameters}) by $D_{\rm expt}$,
figures~\ref{fig:startup_D-4} and~\ref{fig:startup_D-9} are for
$D=D_{\rm expt}$ (type I system) and $D=10^{-5}D_{\rm expt}$ (type II
system) respectively. The overall features of these dispersion
relations are the same as for their time-independent counterparts
(Figs.~\ref{fig:dispersion_phi0.11}
and~\ref{fig:dispersion_coupled_phi0.11_gdot2.0_and_4.0_2D}). In
particular there is, at any time, a well defined peak $k=k_{\rm
  peak}(t)$. This peak shifts along the $k$ axis in time.  At $t=t_0$,
when $\omega^*=0$ by definition, we numerically observe that $k_{\rm
  peak}=0$.  However $k_{\rm peak}$ very quickly attains a value $k^*$
that is (practically) time-independent and well approximated by
Eqn.~\ref{eqn:dispersion_peak} . In this way, the time-dependence of
$k_{\rm peak}$ only occurs at early times $t\gae t_0$, for which the
growth rate is insignificantly small.  We argue, therefore, that we
can choose the ultimate $k_{\rm peak}=k^*$ as the representative
wavevector for the instability.

The time dependent eigenvector at this selected wavevector $k^*$ is
also shown in figures~\ref{fig:startup_D-4} to~\ref{fig:startup_D-9}.
As noted above, the eigenvector encodes the extent to which separation
occurs in each of different order parameters.  Experimentally,
polarised light scattering is sensitive to fluctuations in the
micellar strain, while unpolarised light scattering measures
fluctuations in the overall micellar concentration. In a forthcoming
paper~\cite{upcoming_plane}, we explicitly calculate the unstable
startup static structure factor separately for polarised and
unpolarised light scattering. In this paper, we restrict ourselves to
the overall features on the instability that are deducible from the
eigenvector.

For type I systems at all (unstable) shear
rates (figures~\ref{fig:startup_D-4:b},~\ref{fig:startup_D-4:d}
and~\ref{fig:startup_D-4:f}), and for type II systems at shear rates
that are not too small (figures~\ref{fig:startup_D-9:d}
and~\ref{fig:startup_D-9:f}), the eigenvector is dominated by the flow
variables $\delta\gdot$ and $\delta\tens{W}$ as expected. In contrast,
for the type II system at low shear rates
(Fig.~\ref{fig:startup_D-9:b}) the eigenvector is dominated by
concentration: here the instability is essentially CH demixing,
triggered by flow. At higher shear rates, even in this type II system,
the instability is basically mechanical.  Note finally that
concentration coupling affects the relative contributions of the shear
($W_{xy}$) and normal ($Z$) stresses. Recall that in the uncoupled
limit, $\delta Z\gg \delta W_{xy}$. In this coupled case, the shear
component can be comparable to $\delta Z$ (for moderate applied shear
rate $\bar{\gdot}$) or even larger than $\delta Z$ (low
$\bar{\gdot}$).

\section{Conclusion}
\label{sec:conclusion}

In this paper, we have studied the early-time kinetics of the shear banding
instability in startup flows. Motivated by recent rheo-optical
experiments\cite{DecLerBer01} in which enhanced concentration
fluctuations were observed in the flow/flow-gradient and
flow/vorticity planes at the onset of instability, we performed a
linear stability analysis for coupled fluctuations in shear rate,
viscoelastic stress and concentration using the non-local
Johnson-Segalman model and a two-fluid approach to concentration
fluctuations.

We considered two flow histories. The first assumed an initial
homogeneous state on the intrinsic constitutive curve. Using this, we
defined the spinodal boundaries of the unstable region for slow shear
rate sweeps. (Any real system can in practice shear band via
metastable kinetics before the unstable region is reached. The
analogous ambiguity occurs in the CH calculation for fluid-fluid
demixing.)  For the uncoupled limit of infinite drag between micelles
and solvent at finite collective micellar diffusion constant $D$,
fluctuations in the mechanical variables (shear rate and stress)
decouple from concentration and are unstable when the intrinsic
constitutive curve has negative slope, as expected; the concentration
has its own CH instability for $D<0$.

For finite drag, fluctuations in the mechanical variables couple to
those in concentration via the positive feedback mechanism of Helfand
and Fredrickson, and the unstable region of what was the purely
mechanical instability is broadened. In rapid upward stress sweep
experiments, therefore, ``top'' jumping should in fact occur 
{\em before} the maxiumum in the intrinsic constitutive curve is reached.
For given values of the coupling parameters, the degree of broadening
increases with proximity to an underlying (zero shear) CH demixing
instability.  Accordingly, we classified systems into two types. Type
I systems are far from a CH instability, and the mechanical
instability is only slightly perturbed by concentration coupling. Type
II systems are close to a CH instability.  Type I systems, and type II
systems at high shear rates, show instabilities that are predominantly
mechanical (type A). Type II systems at low shear rates should show a
CH instability perturbed by coupling to flow (type B).

We then discussed the dispersion relations for fluctuations about the
unstable intrinsic constitutive curve.  In the uncoupled limit, there
is a broad plateau in the dispersion relation, with no selected length
scale.  Concentration coupling enhances the instability at short
wavelengths thereby selecting a wavelength. However the typical growth
rates predicted by these dispersion relations are larger than the rate
at which the system can realistically be prepared on the intrinsic
constitutive curve. We therefore explicitly considered shear startup
quenches into the unstable region.  We assumed that the startup flow
can be decomposed into a homogeneous background state, evolving
towards the intrinsic constitutive curve, with small fluctuations that
(consistently with the above remarks) in general go unstable before
the the intrinsic constitutive curve can be attained.  The main
features of the time independent dispersion relations for fluctuations
about the intrinsic constitutive curve are nonetheless preserved: the
mechanical instability shows length scale selection only when coupled
to concentration. Our results for the time dependent eigenvector at
this selected length scale confirmed our classification of
instabilities into types A and B.

In the coupled model, for small values of the diffusion coefficient a
second lobe of instability develops at high shear rates.  This could
clearly have dramatic consequences for any putative coexistence of low
shear and high shear bands, since the high shear band could itself be
unstable. Indeed, the high shear band is often seen to fluctuate
strongly~\cite{BritCall99}, or to break into smaller
bands~\cite{lerouge_note}.  However, in our model this high-shear
instability is highly sensitive to choice of model parameters, and
could be an unrealistic feature.

Our study was confined to fluctuations in the flow gradient direction, and
to the qualitative features of the instability that can be gleaned
from the time-dependent dispersion relations and eigenvectors. In a
future paper, we will present results for the time-dependent unstable
static structure factor in startup, for fluctuations in the entire
flow/flow-gradient plane~\cite{upcoming_plane}.

{\bf Acknowledgments} We thank Paul Callaghan, Ron Larson, Sandra
Lerouge, and Tom McLeish for interesting discussions and EPSRC GR/N
11735 for financial support; this work was supported in part by the
National Science Foundation under Grant No.PHY99-07949.



\begin{thebibliography}{63}
\bibitem[{neg({\natexlab{a}})}]{neglect_inertia}
\bibinfo{note}{The criterion for neglecting the relative inertial terms when
  considering a fluctuation of rate $\omega$ and wavevector $k$ about a state
  of uniform shear rate $\gdot$ is $\mbox{max}[\rho\omega,\rho\gdot]\ll
  \mbox{max}[\eta k^2, \zeta]$.}
\bibitem[{vec()}]{vector_caution}
\bibinfo{note}{Before rescaling, $\bar{\gdot}$, $\bar{\tens{W}}$ and
  $\bar{\phi}$ had dimensions $[T^{-1}]$, $[1]$, and $[1]$. Correspondingly,
  the size and direction of the vector $\vect{u}$ would have been different in
  different units, and any vector operations must be applied with caution.}
\bibitem[{neg({\natexlab{b}})}]{neglect_Y}
\bibinfo{note}{Matrix~\ref{eqn:fluct_in_y_matrix_conc} neglects $\delta Y$,
  which amounts to assuming that the velocity convecting the micellar strain is
  $\vect{v}=\phi\vm+(1-\phi)\vs$ and not $\vm$. It also neglects the positive
  feedback between flow and concentration involving the elastic contribution
  $F^{\rm e}$ to the free energy in Eqn.~\ref{eqn:concentration}.}
\bibitem[{mul()}]{multi_branch}
\bibinfo{note}{There are rare cases of multiple positive branches but never,
  for any regime investigated, in the relevant window of time $t<\tau_{\rm
  inst}$.}
\end{thebibliography}

\end{document}